\documentclass[twocolumn,trackchanges]{aastex62}

\usepackage[utf8]{inputenc}
\usepackage{xcolor}
\usepackage{hyperref}
\usepackage{amsmath,units}
\usepackage{xspace}

\definecolor{chmagenta}{rgb}{0.54, 0.17, 0.88}
\definecolor{placeholder}{rgb}{0.10, 0.50, 0.10}

\newcommand{\oneG}{\textsf{1G}\xspace{}}
\newcommand{\twoG}{\textsf{2G}\xspace{}}
\newcommand{\firstgen}{\textsf{1G+1G}\xspace{}}
\newcommand{\halfgen}{\textsf{1G+2G}\xspace{}}
\newcommand{\secondgen}{\textsf{2G+2G}\xspace{}}

\newcommand{\Beta}{\ensuremath{\mathrm{B}}}
\newcommand{\pret}{\ensuremath{P_\mathrm{ret}}}
\newcommand{\fret}{\ensuremath{F_\mathrm{ret}}}

\newcommand{\CIERA}{Center for Interdisciplinary Exploration and Research in Astrophysics (CIERA), Department of Physics and Astronomy, Northwestern University, 1800 Sherman Avenue, Evanston, IL 60201, USA}
\newcommand{\Monash}{School of Physics and Astronomy, Monash University, VIC 3800, Australia}
\newcommand{\OzGrav}{OzGrav: The ARC Centre of Excellence for Gravitational-Wave Discovery, Clayton, VIC 3800, Australia}
\newcommand{\Caltech}{LIGO, California Institute of Technology, Pasadena, CA 91125, USA}

\newcommand{\change}[1]{{#1}}

\def\GWTCOneGaussalphaLow{\ensuremath{0.83}}

\def\GWTCOneGaussmmaxMed{\ensuremath{47.5}}

\def\GWTCOneGaussmmaxMinus{\ensuremath{13.5}}
\def\GWTCOneGaussmmaxPlus{\ensuremath{16.5}}

\def\GWTCOneGaussmppLow{\ensuremath{22}}
\def\GWTCOneGaussmppMed{\ensuremath{31}}
\def\GWTCOneGaussmppHigh{\ensuremath{38}}
\def\GWTCOneGaussmppMinus{\ensuremath{8.6}}
\def\GWTCOneGaussmppPlus{\ensuremath{7.1}}

\def\GWTCOneGaussdeltachiULNinetyNine{\ensuremath{0.32}}
\def\GWTCOneGaussdeltachiULNinety{\ensuremath{0.16}}

\def\GWTCOneGaussBranchingHalfMed{\ensuremath{2.5\times 10^{-3}}}

\def\GWTCOneGaussBranchingHalfULNinetyNine{\ensuremath{0.049}}

\def\GWTCOneGaussBranchingTwoMed{\ensuremath{3.1\times 10^{-6}}}

\def\GWTCOneGaussBranchingTwoULNinetyNine{\ensuremath{1.2\times 10^{-3}}}
\def\GWTCOneGaussOneGSpinULNinety{\ensuremath{0.57}}
\def\GWTCOneGaussOneGMassULNinetyNine{\ensuremath{44}}
\def\GWTCOneGaussAllMassULNinetyNine{\ensuremath{45}}

\def\GWTCOneNoZeroSpinGaussBranchingHalfMed{\ensuremath{8.1\times 10^{-4}}}

\def\GWTCOneNoZeroSpinGaussBranchingHalfULNinetyNine{\ensuremath{0.018}}

\def\GWTCOneNoZeroSpinGaussBranchingTwoMed{\ensuremath{3.3\times 10^{-7}}}

\def\GWTCOneNoZeroSpinGaussBranchingTwoULNinetyNine{\ensuremath{1.6\times 10^{-4}}}
\def\GWTCOneNoZeroSpinGaussOneGSpinULNinety{\ensuremath{0.54}}

\begin{document}

\title{Black hole genealogy: Identifying hierarchical mergers with gravitational waves}
\shorttitle{Black hole genealogy}
\shortauthors{Kimball \textit{et al}.}

\author[0000-0001-9879-6884]{Chase~Kimball}
\correspondingauthor{Chase~Kimball}
\email{CharlesKimball2022@u.northwestern.edu}
\affiliation{\CIERA}

\author[0000-0003-2053-5582]{Colm~Talbot}
\affiliation{\Caltech}
\affiliation{\Monash}
\affiliation{\OzGrav}

\author[0000-0003-3870-7215]{Christopher~P~L~Berry}
\affiliation{\CIERA}

\author[0000-0003-4957-6679]{Matthew~Carney}
\affiliation{\CIERA}

\author[0000-0002-0147-0835]{Michael~Zevin}
\affiliation{\CIERA}

\author[0000-0002-4418-3895]{Eric~Thrane}
\affiliation{\Monash}
\affiliation{\OzGrav}

\author[0000-0001-9236-5469]{Vicky~Kalogera}
\affiliation{\CIERA}

\begin{abstract}
In dense stellar environments, the merger products of binary black hole mergers may undergo additional mergers. These hierarchical mergers are naturally expected to have higher masses than the first generation of black holes made from stars.
The components of hierarchical mergers are expected to have significant characteristic spins, imprinted by the orbital angular momentum of the previous mergers. However, since the population properties of first-generation black holes are uncertain, it is difficult to know if any given merger is first-generation or hierarchical. We use observations of gravitational waves to reconstruct the binary black hole mass and spin spectrum of a population \change{including the possibility of hierarchical mergers}. We employ a phenomenological model that captures the properties of merging binary black holes from simulations of \change{globular clusters}. Inspired by recent work on the formation of low-spin black holes, we include a zero-spin subpopulation. We analyze binary black holes from LIGO and Virgo's first two observing runs, and find that this catalog is consistent with having no hierarchical mergers. We find that the most massive system in this catalog, GW170729, is mostly likely a first-generation merger, having a $4\%$ probability of being a hierarchical merger assuming a $5 \times 10^5 M_{\odot}$ globular cluster mass. Using our model, we find that $99\%$ of first-generation black holes in coalescing binaries have masses below \GWTCOneGaussOneGMassULNinetyNine{}  $M_{\odot}$, and the fraction of binaries with near-zero component spins is \change{less than \GWTCOneGaussdeltachiULNinety{} ($90\%$ probability).} Upcoming observations will determine if hierarchical mergers are a common source of gravitational waves.
\end{abstract}

\keywords{
Gravitational wave sources  ---  Gravitational wave astronomy --- Astrophysical black holes  ---  Hierarchical models
}

\section{Introduction}\label{intro}

The gravitational-wave (GW) observations of LIGO \citep{TheLIGOScientific:2014jea} and Virgo \citep{TheVirgo:2014hva} have revealed a population of stellar-mass binary black holes \citep{Abbott:2016blz,LIGOScientific:2018mvr,LIGOScientific:2020stg}. 
These black holes range in mass over $\sim 7$--$50 M_\odot$, extending beyond the masses observed in X-ray binaries \citep{TheLIGOScientific:2016htt,Miller:2014aaa}. 
Since black hole systems can encode information about how their progenitor systems evolve \citep{TheLIGOScientific:2016htt,Abbott:2017vtc,Mandel:2018hfr}, this new population of black holes observed via GWs has broadened our understanding of the physical processes that shape the mass spectrum of stellar-origin black holes. 
Already, GW observations hint at a dearth of stellar-mass black holes with component masses $\gtrsim 45 M_\odot$ \citep{LIGOScientific:2018jsj}, as predicted by theory decades ago.

Black holes are the end point of stellar evolution for stars $\gtrsim 20 M_\odot$ \citep{Woosley:2002zz}. 
Though more massive stars typically result in more massive black holes, the mapping between initial stellar mass and remnant mass is affected by many physical processes including stellar winds, stellar rotation, and binary interactions \citep{Belczynski:2010tb,Spera:2015vkd,Kruckow:2018slo,Neijssel:2019irh,Ertl:2019zks}. 
Additionally, stellar evolution does not predict a simple continuum that persists to arbitrarily high black hole masses. 
When stellar cores reach $\sim 50 M_\odot$ they become become susceptible to pair instability \citep{Fowler:1964zz}. 
In this process, high-energy photons undergo electron--positron pair production, causing a drop in the radiation pressure supporting the stellar core. 
The core subsequently contracts, increasing the temperature, triggering nuclear burning of carbon, oxygen, and silicon \citep[][]{Woosley:2014lua}. 
Stellar cores of $\sim 35$--$65 M_\odot$ undergo pulsational pair instabilities \citep[PPSNe;][]{Woosley:2007qp,Woosley:2016hmi,Marchant:2018kun}, where the star sheds large amounts of mass prior to collapse, limiting the resultant mass of the remnant black hole. 
Stars with cores of $\sim 65$--$135 M_\odot$ are subject to pair-instability supernovae \citep[PISNe;][]{Barkat:1967zz,Fryer:2000my,Heger:2001cd}, where the instability results in the complete disruption of the star and no remnant black hole. 
Stellar evolution theory predicts a gap in the black hole mass spectrum between $\approx 45$--$135 M_\odot$ \citep{Belczynski:2016jno,Spera:2017fyx,Stevenson:2019rcw}. 

Measuring the bounds of the PISN mass gap will provide insights into stellar evolution and fundamental physics \citep{Farr:2019twy,Farmer:2019jed,vanSon:2020zbk,Talbot:2018cva}. 
However, one needs to account for the dynamical processes that can lead to black holes in this mass range. 
In dense stellar environments, such as globular clusters and nuclear star clusters, gravitational encounters of black holes in the cluster core harden the orbits of binary black hole systems, facilitating mergers within the cluster \citep[e.g.,][]{Heggie:1975tg,Banerjee:2009hs,Rodriguez:2016kxx}. 

If these merger products remain in the cluster environment, they can potentially merge again. 
These hierarchical mergers are characterized by a higher masses and spins than is typical of black holes born from stars \citep{Miller:2002pg,Gerosa:2017kvu,Fishbach:2017dwv,Kimball:2019mfs,Sedda:2020vwo,Baibhav:2020xdf}. 
Dense stellar environments are prime locations for facilitating such hierarchical mergers, which exhibit unique intrinsic properties that can be measured with GWs.

Identifying black holes formed through previous mergers requires knowledge of the initial mass spectrum of black holes formed through direct stellar collapse \citep{Kimball:2019mfs,Doctor:2019ruh}.
Given the uncertainties in massive star evolution and binary stellar evolution, the properties of the natal black hole population are uncertain---it is something we aim to determine from GW observations. 
Therefore, it is essential to simultaneously infer the properties of the natal black hole population as part of our hierarchical mergers model.
By doing so we can reconstruct valuable information about the origins of binary black holes. 
For example, the mass spectrum of the natal black hole population contains information on the stellar mass-loss rates \citep{Stevenson:2015bqa,Barrett:2017fcw}. 
Meanwhile, the fraction of merger products  that go on to merge again encodes information on the physics of dense stellar environments. 
Only a fraction of black holes formed from binary black hole mergers are retained within a cluster, since the merger product receives a recoil kick from the anisotropic GW emission \citep{Blanchet:2013haa,Campanelli:2007ew,Lousto:2011kp,Sperhake:2014wpa} or can be subsequently ejected through close dynamical interactions with other objects \citep{Heggie:1975tg,PortegiesZwart:1999nm,Moody:2008ht,Downing:2010hq}. 
The fraction retained depends on the mass and size of the cluster, and crucially upon the spins of the progenitor black holes \citep{Rodriguez:2017pec,Rodriguez:2019huv,Gerosa:2019zmo,Banerjee:2020bwk}. 
Furthermore, the number of hierarchical mergers may enable us to determine the dominant formation channel for binary black holes.

In this study, we investigate how hierarchical binary black hole mergers can be identified within a population of GW observations. 
We focus on formation in globular clusters, where, due to the shallow gravitational potential, merger products typically cannot proceed through more than one additional merger before being ejected. 
We refer to the population of black holes formed from standard stellar evolution as first generation (\oneG), and black holes that result from a binary black hole merger of \oneG{} components as second generation (\twoG). 
Hierarchical mergers, involving one or more \oneG{} black hole, are denoted \halfgen{} and \secondgen{} depending on whether the merger contains one or two \twoG{} black holes.
First-generation mergers are denoted \firstgen.

Using simple phenomenological models for the properties of \firstgen, \halfgen, and \secondgen{} binaries, we perform hierarchical inference to determine the properties and rates of these different subpopulations. 
These phenomenological models are a natural extension of previous studies of the mass and spin distributions of binary black holes \citep{Fishbach:2017zga,Talbot:2018cva,Wysocki:2018mpo,LIGOScientific:2018jsj} and are explained in Sec.~\ref{sec:models}. 
The hierarchical inference methodology using these models is explained in Sec.~\ref{sec:inference}. 
We apply our methodology to the set of binary black holes presented in GWTC-1 \citep{LIGOScientific:2018mvr} in Sec.~\ref{sec:GWTC-1}, \change{and discuss the inferred population hyperparameters in Appendix~\ref{sec:appendix}}. 
In Appendix~\ref{sec:GW190412}, we consider how results change upon adding GW190412 \citep{LIGOScientific:2020stg} to the GWTC-1 population.
In future work, we will extend this analysis to events found by external searches \citep{Nitz:2019hdf, Venumadhav:2019tad, Zackay:2019tzo,Venumadhav:2019lyq,Zackay:2019btq}. 
We find that observations are consistent with all binaries being \firstgen{} \citep{Kimball:2019mfs,Chatziioannou:2019dsz,Yang:2019cbr}; 
however, if we include the possibility that some \oneG{} black holes are born with near-zero spins \citep{Qin:2018nuz,Fuller:2019sxi,Belczynski:2017gds}, we find a small probability of GW170729 containing a \twoG{} black hole \change{using our models for globular clusters}. 
Our conclusions are summarized in Sec.~\ref{sec:conclusions}.

\section{Population model}\label{sec:models}

Phenomenological models are computationally efficient tools for parameterizing black hole population properties. 
The model we develop in this study approximates the detectable population of merging binary black holes from globular clusters, and is designed to capture the main features of binaries formed through hierarchical mergers.
The model is constructed using \change{population hyperparameters} $\boldsymbol{\Lambda}$ that describe the \firstgen{} black hole population. 

We assume that the overall population of binary black holes consists of three subpopulations:  \firstgen, \halfgen{} and \secondgen{} binaries.
We neglect the probability of higher-order mergers (containing a $\geq$ \twoG{} component) in this analysis since the number of these mergers is negligible in globular cluster models \citep{Rodriguez:2019huv,Sedda:2020vwo}. 
However, dense stellar environments such as those in galactic nuclei, nuclear star clusters \citep{Antonini:2018auk}, and active galactic nucleus disks \citep{Yang:2019cbr}, may retain higher-order merger products and our approach can be expanded to include their contribution.

The fractions of total mergers associated with each generation are denoted $\zeta_\firstgen(\boldsymbol{\Lambda})$, $\zeta_\halfgen(\boldsymbol{\Lambda})$ and $\zeta_\secondgen(\boldsymbol{\Lambda})$.
Since only a small fraction of \twoG{} black holes are retained in the fiducial cluster and able to form a new binary, we expect that $\zeta_\firstgen(\boldsymbol{\Lambda}) \gg \zeta_\halfgen(\boldsymbol{\Lambda}) \gg \zeta_\secondgen(\boldsymbol{\Lambda})$. 
By unitarity, we have
\begin{equation}\label{eq:unitarity}
    \zeta_\firstgen(\boldsymbol{\Lambda}) + \zeta_\halfgen(\boldsymbol{\Lambda}) + \zeta_\secondgen(\boldsymbol{\Lambda}) = 1 .
\end{equation}
The fraction of binaries in each subpopulation depends upon the population properties of the \firstgen{} binary black holes. 
In particular, the distributions of component spins and mass ratio have a strong effect on the recoil kick during merger.

For each generation, we define an astrophysically motivated prior on the properties $\boldsymbol{\theta}$ describing individual binary black holes, such as their masses and spins. 
We decompose the overall prior for a given generation into priors on the primary mass $m_1$, mass ratio $q = m_2/m_1$, spin magnitudes $\chi_1$ and $\chi_2$, spin orientations ${z_1}\equiv\cos\theta_1$ and ${z_2}\equiv\cos\theta_2$ (where $\theta_i$ is the angle between the black hole spin and the orbital angular momentum vector), and extrinsic parameters ${\vartheta}$. 
The prior on the extrinsic parameters is assumed to be the same for all generations: mergers are uniformly distributed in comoving volume and we employ standard priors for other extrinsic parameters. 

The population model is described in the following subsections.
In Sec.~\ref{firstgen}, we describe a model for the mass and spin distributions of \firstgen{} binary black holes \citep{Wysocki:2018mpo,Talbot:2018cva,Talbot:2017yur,LIGOScientific:2018jsj}. 
The population of \firstgen{} binary black holes forms the cornerstone of our models, and the properties of merger products are set based upon this. 
In Sec.~\ref{NextGen}, we describe our prescription to estimate the mass and spin distributions of \halfgen{} and \secondgen{} binaries given the \firstgen{} distribution. 
In Sec.~\ref{branching}, we describe our method for calculating the generational fractions $\zeta_\firstgen$, $\zeta_\halfgen$, and $\zeta_\secondgen$ given our population model. 
The hierarchical inference method we outline in Sec.~\ref{sec:inference} can be adapted to use alternative phenomenological models as improved descriptions are developed. 
The phenomenological method presented here predicts distributions that are qualitatively similar to simulations of globular clusters \citep[e.g.,][]{Rodriguez:2019huv}. 

\subsection{\firstgen{} binaries}\label{firstgen}

\subsubsection{Primary mass}

Following \citet{LIGOScientific:2018jsj}, we model the distribution of \firstgen{} black hole primary mass $m_1$ using the prescription from \citet{Talbot:2018cva}
\begin{align}
  \pi(m_1|&\alpha,m_\mathrm{min},m_\mathrm{max},\lambda_m,\mu_m,\sigma_m,\firstgen) = \nonumber\\
    & \left[ (1-\lambda_m) \mathcal{A} m_1^\alpha \Theta(m_\mathrm{max}-m_1) \right. + \nonumber\\
    & \left. \lambda_m \mathcal{B} N(m_1 |\mu_m, \sigma_m) \right] ,
    \label{eq:m1-model}
\end{align}
where $\left\{\alpha,m_\mathrm{min},m_\mathrm{max},\lambda_m,\mu_m,\sigma_m,\right\} \in \boldsymbol{\Lambda}$ are the \change{population hyperparameters} defining this distribution. 
This model includes two components. 
The first is a truncated power-law distribution with spectral index $\alpha$ and a maximum mass of $m_\mathrm{max}$ (enforced by the Heaviside step function $\Theta$).
The second is a Gaussian component with mean $\mu_m$ and standard deviation $\sigma_m$.
The parameter $\lambda_m$ is a mixing fraction, which determines the fraction of binaries associated with either component. 
The factors $\mathcal{A}$ and $\mathcal{B}$ are normalization constants that depend on the other \change{population hyperparameters}.
This mass distribution is chosen to enforce the expected cutoff in the black hole mass spectrum from PISNe \citep{Heger:2002by,Belczynski:2016jno,Fishbach:2017zga}, with the Gaussian capturing a buildup from PPSNe \citep{Woosley:2016hmi,Marchant:2018kun,Talbot:2018cva}.

\subsubsection{Mass ratio} 

Following \citet{LIGOScientific:2018jsj}, we model the \firstgen{} mass ratio $q$ using a power-law distribution \citep{Talbot:2018cva},
\begin{align}
    \pi(q|&m_1,\beta_q,m_\mathrm{min},\firstgen)=\mathcal{C}(m_1) m_2^{\beta_q} \Theta(m_1-m_2),
    \label{eq:q-model}
\end{align}
defined using \change{population hyperparameters} $\{m_1,\beta_q,m_\mathrm{min}\} \in \boldsymbol{\Lambda}$. 
Here $\beta_q$ is the power-law index, and $\mathcal{C}$ is a normalization constant.

\subsubsection{Spin magnitudes}

We assume that the spin magnitudes of both black holes $\chi_1$ and $\chi_2$ are described by the same distribution,
\begin{align}
    \pi(\chi| \lambda_0, \alpha_\chi, \beta_\chi, \firstgen) = \lambda_0 \delta(\chi) + (1-\lambda_0) \Beta(\chi | \alpha_\chi, \beta_\chi),
    \label{eq:chi-model}
\end{align}
described by \change{population hyperparameters} $\{\lambda_0, \alpha_\chi, \beta_\chi \} \in \boldsymbol{\Lambda}$
Here, $\Beta$ is a Beta distribution parameterized by shape parameters $\alpha_\chi$ and $\beta_\chi$ \citep{Wysocki:2018mpo}.

However, a simple Beta distribution will struggle to capture the morphology of the true population if a significant fraction of binary black holes have low ($\lesssim 0.01$) natal spins, which is anticipated to be the case if angular momentum transport in massive stars is efficient \citep{Qin:2018nuz,Fuller:2019sxi}.
The mixing parameter $\lambda_0$ controls the fraction of black holes merging with negligible spin.
We assume that the spin of the primary black hole in a binary is independent from the spin of the secondary black hole.

\subsubsection{Spin orientation} 

The orientation of black hole spin can be parameterized using the cosine of the polar angle between the spin vector and the Newtonian orbital angular momentum $z_i = \cos\theta_i$.
In \citet{LIGOScientific:2018jsj}, the orientation of black hole spin was modeled using a mixture model \citep{Talbot:2017yur}
\begin{align}
    \pi(z_1, z_2|& \zeta_\mathrm{iso}, \sigma_1, \sigma_2, \firstgen) = \zeta_\mathrm{iso} U(z_1)U(z_2) \nonumber\\ 
    & + (1-\zeta_\mathrm{iso}) N_\mathrm{t}(z_1|0,\sigma_1) 
    N_\mathrm{t}(z_2|0,\sigma_2) ,
    \label{eq:iso-spin}
\end{align}
defined with \change{population hyperparameters} $\{\zeta_\mathrm{iso}, \sigma_1, \sigma_2\} \in \boldsymbol{\Lambda}$. 
Here $\zeta_\mathrm{iso}$ is the fraction of binaries that are drawn from a distribution with isotropic spin orientations (uniform in $z_1$ and $z_2$).
The isotropic distribution is expected for dynamically assembled binaries because the stellar progenitors did not coevolve. 
Binaries that are not drawn from this uniform distribution $U$ are drawn from a truncated normal distribution $N_\mathrm{t}$.
The normal distribution is centered on $z=0$ corresponding to aligned spin with width determined by the standard deviations $\sigma_1$ and $\sigma_2$.
The truncated normal distribution represents the binaries formed in the galactic field, where spins are predicted to be generally aligned, with some scatter due to supernova kicks \citep{Rodriguez:2016vmx}. 
For this analysis, we set $\zeta_\mathrm{iso}=1$, which effectively adopts the framework that all binaries are dynamical mergers:
\begin{align}
    \pi(z_1, z_2|\sigma_1, \sigma_2, \firstgen) = U(z_1)U(z_2) .
\end{align}
For future work, this model could be extended to reintroduce $\zeta_\mathrm{iso}$ and only to consider hierarchical mergers from the fraction of events formed dynamically.

\subsection{\halfgen{} and \secondgen{} binaries}\label{NextGen}

\subsubsection{Primary mass}

Our model for the primary mass distributions for \halfgen{} and \secondgen{} mergers is built on the premise that \secondgen{} black holes are roughly twice as massive as \firstgen{} black holes.\footnote{While mass energy is radiated away in GWs so that the remnant mass is a few percent less than the sum of the primary and secondary masses \citep{Reisswig:2009vc,Healy:2014yta,Jimenez-Forteza:2016oae}, this is negligible compared to astrophysical modeling uncertainties.} 
We make the simplifying assumption that in a \halfgen{} binary, the primary is always the \twoG{} black hole \citep{Kimball:2019mfs}. 
Thus, the \halfgen{} and \secondgen{} primary mass spectra are modeled as
\begin{align}
    \pi(m_1|\boldsymbol{\Lambda}, \halfgen) \propto & {}\: \pi\left(\frac{m_1}{2}\middle|\boldsymbol{\Lambda}, \firstgen\right), \\
    \pi(m_1|\boldsymbol{\Lambda}, \secondgen) \propto & {}\: \pi\left(\frac{m_1}{2}\middle|\boldsymbol{\Lambda}, \firstgen\right),
\end{align}
This representation is found to qualitatively match the results of globular cluster simulations \citep[e.g.,][]{Rodriguez:2017pec,Rodriguez:2019huv}.

\subsubsection{Mass ratio}

Since we expect that \halfgen{} and \secondgen{} binaries are formed dynamically, the mass ratio distributions should depend upon mass segregation and the dynamical interactions that form binaries inside dense stellar environments. 
We calibrate our mass ratio distributions against the results of globular cluster simulations from \citet{Rodriguez:2019huv}.
For \halfgen{} binaries, we adopt a model where the mass ratio distribution peaks around $q \sim 0.5$. 
We find that the distribution recovered from cluster simulations may be approximated as
\begin{align}
    \pi(q|\boldsymbol{\Lambda}, \halfgen) \propto &  
        \begin{cases}
           \pi(q|\boldsymbol{\Lambda}, \firstgen)^{1.5} & q\leq 1/2 \\ 
           \pi(1 - q|\boldsymbol{\Lambda}, \firstgen)^{-1.5} & q> 1/2
        \end{cases}.
\end{align}
An alternative parameterization, producing a similar form, is given in \citet{Chatziioannou:2019dsz}. 
The most important feature of the \halfgen{} distribution is that it peaks away from $q = 1$, as this distinguishes it from the \firstgen{} and \secondgen{} distributions.

For \secondgen{} binaries we find that
\begin{align}
    \pi(q|\boldsymbol{\Lambda}, \secondgen) \propto & {} \pi(q|\boldsymbol{\Lambda},\firstgen)^4 
\end{align}
produces qualitative agreement with predictions from \citet{Rodriguez:2019huv}. 
This distribution is more tightly peaked at $q \sim 1$ than the \firstgen{} population, reflecting the preference for dynamically formed binary mergers to by dominated by the most massive components in the cluster \citep{1996ApJ...467..359H,1993ApJ...415..631S,Downing:2010hq}.

\subsubsection{Spins}

The spin magnitude of post-merger remnants is primarily determined by the orbital angular momentum of the progenitor binary \citep{Pretorius:2005gq,Buonanno:2007sv,Gonzalez:2006md}.
For typical binaries (with mass ratio $q \approx 1$ and low spins) the remnant spin is $\approx0.67$.
We therefore adopt for \halfgen{} spins
\begin{align}
    \pi(\chi_1|\boldsymbol{\Lambda},\halfgen) = & {} \: \Beta(\chi_1|14.14,6.97), \\
    \pi(\chi_2|\halfgen) = & {} \: \pi(\chi_2|\boldsymbol{\Lambda},\firstgen) ,
\end{align}
and for \secondgen{} spins
\begin{align}
    \pi(\chi_1|\boldsymbol{\Lambda},\secondgen) = & {} \: \Beta(\chi_1|14.14,6.97) , \\
    \pi(\chi_2|\boldsymbol{\Lambda},\secondgen) = & {} \: \Beta(\chi_2|14.14,6.97) .
\end{align}
Here, $\Beta(\mu,\sigma)$ is a beta function with shape parameters $\alpha_{\chi}$, $\beta_{\chi}$, which corresponds to a mean $0.67$ and standard deviation $0.1$ \citep{Fishbach:2017dwv,Chatziioannou:2019dsz,Sedda:2020vwo}.
We assume that the \halfgen{} and \secondgen{} spins are isotropically oriented, the same as for \firstgen{} binaries.

\subsection{Retention fraction}\label{branching}

Given a \firstgen{} population, the branching ratios of the \halfgen{} and \secondgen{} populations are determined by the fraction of \firstgen{} merger products that are retained in the cluster. 
During the coalescence of a binary black hole, the anisotropic emission of GWs imparts a kick on the remnant. 
The magnitude of the kicks depends sensitively on the spin and mass ratio of the binary \citep{Gonzalez:2006md,Campanelli:2007ew,Brugmann:2007zj,Lousto:2011kp,Varma:2018aht}, and can far exceed the typical escape velocities of globular clusters ($\sim 30$--$50~\mathrm{km\,s}^{-1}$ at $z=0$), ejecting merger products and leaving them unavailable to form new generations of binary black holes \citep{Merritt:2004xa,Moody:2008ht,Varma:2020nbm}. 
Therefore, the branching ratios of the \halfgen{} and \secondgen{} popoulations are sensitive to the distribution of mass ratios and component spins in the \firstgen{} population, as well as the mass and size of the cluster.
 
In order to estimate the retention fraction, we begin by calculating the probability $\pret(\chi_1,\chi_2,q)$ that the remnant of a merging binary with component spins and mass ratio $(\chi_1,\chi_2,q)$ will be retained in a cluster potential following the GW recoil kick.
For our cluster model, we adopt a Plummer potential \citep{Plummer:1911zza} with mass distribution
\begin{equation}
\label{eq:plummermass}
    \rho_p(r) = \frac{3M_\mathrm{c}}{4\pi r_\mathrm{c}^3}\left(1+\frac{r^2}{a_\mathrm{c}^2}\right)^{-5/2}.
\end{equation}
We assume a cluster mass $M_\mathrm{c} = 5 \times 10^5 M_{\odot}$ and a Plummer radius $r_\mathrm{c} = 1~\mathrm{pc}$ to represent a fiducial globular cluster. 
For a given $\{\chi_1, \chi_2, q\}$ we sample merger locations following Eq.~\eqref{eq:plummermass} and sample component spin-tilts isotropically, then calculate recoil velocities according to \citet{Gerosa:2016sys} and check against the local escape velocity to obtain $\pret(\chi_1,\chi_2,q)$. 

\begin{figure}
\includegraphics[width=0.45\textwidth]{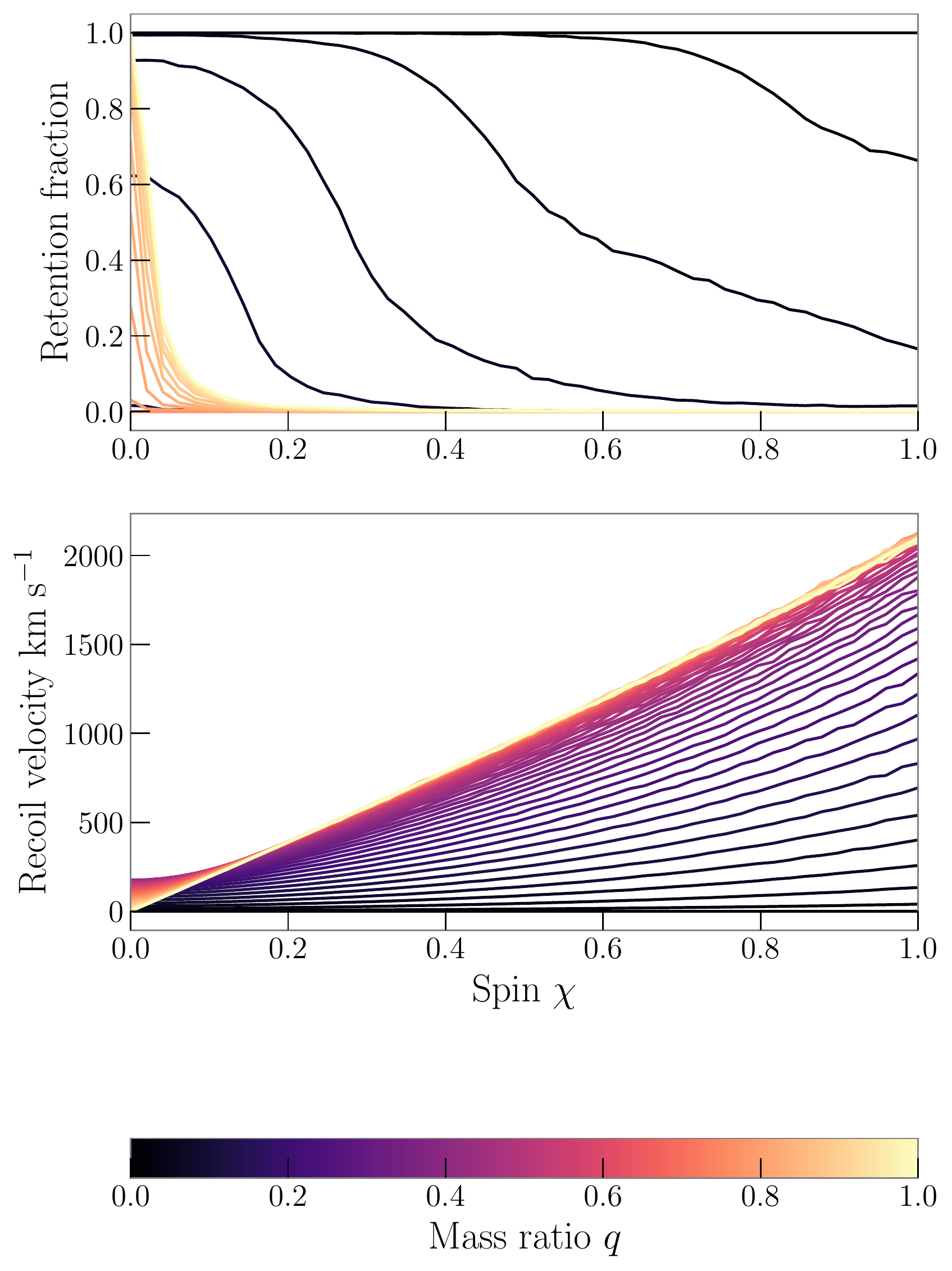}
\caption{
\textit{Top:} the retention fractions $\pret$ assuming a $5\times10^5 M_{\odot}$ cluster with a $1~\mathrm{pc}$ Plummer radius. 
\textit{Bottom:} recoil velocities for equal component-spin binary black holes, colored according to mass ratio $q\equiv m_2/m_1$. For each $\{\chi,q\}$ configuration, we sample spin orientations isotropically and plot the mean recoil velocity. }
\label{fig:retention}
\end{figure}

Figure~\ref{fig:retention} shows $\pret(\chi_1,\chi_2,q)$ for the case of equal spin magnitudes. 
\pret{} is negligible when component spins are $\gtrsim 0.1$, except in the regime of extreme mass ratios ($q \to 0$) where recoil velocities disappear. 
Therefore, nearly all \firstgen{} binaries with appreciable spins will form merger products that are promptly ejected from the fiducial cluster and will be unable to form hierarchical mergers. 
We see that a subpopulation of \oneG{} black holes with negligible spin, represented by the delta function in Eq.~\eqref{eq:chi-model}, is a key ingredient for hierarchical mergers.

For a population determined by \change{population hyperparameters} $\boldsymbol{\Lambda}$, we calculate the fraction \fret{} of \firstgen{} remnants that are retained in our fiducial cluster as 
\begin{align}
    \fret(\boldsymbol{\Lambda}) = & \int \mathrm{d}q \int \mathrm{d} \chi_1 \int \mathrm{d} \chi_2 \, \pi(\chi_1|\boldsymbol{\Lambda},\firstgen) \\\nonumber
    &
    \pi(\chi_2|\boldsymbol{\Lambda},\firstgen) \nonumber
    \pi(q|\boldsymbol{\Lambda},\firstgen) \nonumber \pret(\chi_1,\chi_2,q).
    \label{eq:fret}
\end{align}
Here, $\pi(q|\boldsymbol{\Lambda},\firstgen)$ and $\pi(\chi|\boldsymbol{\Lambda},\firstgen)$ are the \firstgen{} mass ratio and component-spin distributions.

\subsection{Branching ratios}

Using \fret($\boldsymbol{\Lambda}$), we calculate hierarchical branching ratios given a \firstgen{} population with mass and spin distributions determined by \change{population hyperparameters} $\boldsymbol{\Lambda}$. 
Let $R_{\firstgen}$, $R_{\halfgen}$, and $R_{\secondgen}$ be the rates of \firstgen{}, \halfgen{}, and \secondgen{} mergers, respectively, averaged over the lifetime of the cluster. 
The number of \twoG{} black holes available to form new binaries is proportional to $\fret R_\firstgen$. 
Therefore, we expect
\begin{align}
     R_\halfgen = & \: \xi_\halfgen{} \fret(\boldsymbol{\Lambda})  R_\firstgen, \\
     R_\secondgen = & \: \xi_\secondgen{} [\fret(\boldsymbol{\Lambda})]^2  R_\firstgen ,
\end{align}
where the constants of proportionality $\xi_\halfgen{}$ and $\xi_\halfgen{}$ are set by the dynamical processes within the cluster, such as the frequency at which binaries form. 
Based on comparison with simulations \citep{Rodriguez:2019huv}, we find that $\xi_\halfgen{} \simeq  1/2$ and $\xi_\secondgen{} \simeq 1/8$ are good approximations. 
From the rates we can define branching ratios,
\begin{align}
    \Gamma_\halfgen \equiv & \: \frac{R_\halfgen}{R_\firstgen} \propto \fret(\boldsymbol{\Lambda}), \\
     \Gamma_\secondgen \equiv & \: \frac{R_\secondgen}{R_\firstgen} \propto [\fret(\boldsymbol{\Lambda})]^2 
\end{align} 
Since $\fret$ is small, we have $\Gamma_\secondgen \ll \Gamma_\halfgen \ll 1$.

We combine these branching ratios with our individual \firstgen{}, \halfgen{}, and \secondgen{} population distributions to construct a multigenerational mixture model:
\begin{align}
    \pi_\mathrm{hier}({\mathrm{\theta}}|\boldsymbol{\Lambda}) = \,& \: \zeta_{\firstgen}(\boldsymbol{\Lambda})\pi(\boldsymbol{\theta}|\boldsymbol{\Lambda},\firstgen)\nonumber\\
    &+\zeta_{\halfgen}(\boldsymbol{\Lambda})\pi(\boldsymbol{\theta}|\boldsymbol{\Lambda},\halfgen)\nonumber\\
    &+\zeta_{\secondgen}(\boldsymbol{\Lambda})\pi(\boldsymbol{\theta}|\boldsymbol{\Lambda},\secondgen),
\end{align}
where
\begin{align}
    \zeta_{\firstgen} = & \: \frac{1}{1+\Gamma_{\halfgen}+\Gamma_{\secondgen}}, \\
    \zeta_{\halfgen} = & \: \frac{\Gamma_{\halfgen}}{1+\Gamma_{\halfgen}+\Gamma_{\secondgen}}, \\
    \zeta_{\secondgen} = & \: \frac{\Gamma_{\secondgen}}{1+\Gamma_{\halfgen}+\Gamma_{\secondgen}}.
\end{align}
We use the GWTC-1 catalog of GW observations to constrain this model, and infer the \change{population hyperparameters} $\boldsymbol{\Lambda}$, and obtain the odds that any of the observations are from a hierarchical merger.

\section{Population Inference}\label{sec:inference}

Given a set of \change{population hyperparameters} $\boldsymbol{\Lambda}$, the overall likelihood of an observation is

\begin{align}
    \mathcal{L}_\mathrm{hier}(d_i|\boldsymbol{\Lambda}) = \frac{1}{P_{\mathrm{det}}(\boldsymbol{\Lambda})} \int \mathrm{d}\boldsymbol{\theta}
    \, L(d_i|\boldsymbol{\theta}) \pi_\mathrm{hier}(\boldsymbol{\theta}|\boldsymbol{\Lambda}),
    \label{eq:overall-single}
\end{align}
where we use $d_i$ to denote the GW data associated with the $i$-th observation, $L(d_i|\boldsymbol{\theta})$ is the likelihood of the data given the source parameters $\boldsymbol{\theta}$ \citep{Cutler:1994ys,TheLIGOScientific:2016wfe}, $\pi_\mathrm{hier}(\boldsymbol{\theta}|\boldsymbol{\Lambda})$ is the population model defined in Sec.~\ref{sec:models}, and $P_{\mathrm{det}}(\boldsymbol{\Lambda})$ is the fraction of all astrophysical events which are observed and accounts for selection biases \citep{Thrane:2018qnx,Mandel:2018mve}. 
The fraction $P_{\mathrm{det}}(\boldsymbol{\Lambda})$ scales as the surveyed space-time volume $VT(\boldsymbol{\Lambda})$ of the detector network for a binary black hole population with \change{population hyperparameters} $\boldsymbol{\Lambda}$; 
we calculate $VT(\boldsymbol{\Lambda})$ analytically following \citet{Finn:1992xs}, using a single-detector network with a median (over observing times from the first and second observing runs) LIGO Hanford noise curve and signal-to-noise ratio threshold of $8$. 
The overall likelihood in Eq.~\eqref{eq:overall-single} can be broken into pieces associated with each generation,
\begin{align}
    \mathcal{L}_\mathrm{hier}(d_i|\boldsymbol{\Lambda}) = & {} \: \frac{1}{P_{\mathrm{det}}(\boldsymbol{\Lambda})} \left[
    \zeta_\firstgen(\boldsymbol{\Lambda})\mathcal{L}(d_i|\boldsymbol{\Lambda},\firstgen) \right. \nonumber\\ 
    & +\left. {} \zeta_\halfgen(\boldsymbol{\Lambda})\mathcal{L}(d_i|\boldsymbol{\Lambda},\halfgen) \right.
    \nonumber\\
    & +\left. {} \zeta_\secondgen(\boldsymbol{\Lambda})\mathcal{L}(d_i|\boldsymbol{\Lambda},\secondgen) \right],
    \label{eq:overall-single-sub}
\end{align}
where

\begin{align}
    \mathcal{L}(d_i|\boldsymbol{\Lambda},\firstgen) = \int \mathrm{d}\boldsymbol{\theta}
    \, L(d_i|\boldsymbol{\theta}) \pi(\boldsymbol{\theta}|\boldsymbol{\Lambda},\firstgen),
\end{align}
and likelihoods for the other generations are defined similarly.

For a set of $N$ detections (described by data $\vec{d}$), the total likelihood becomes
\begin{align}
\label{eq:Likelihood}
    \mathcal{L}_\mathrm{tot}(\vec{d}|\boldsymbol{\Lambda}) = & \prod_i^N \mathcal{L}_\mathrm{hier}(d_i|\boldsymbol{\Lambda}).
\end{align}  
To calculate the total likelihood, we use samples drawn from the black hole parameter posterior probability distributions
\begin{equation}
p(\boldsymbol{\theta}|d_i) = \frac{L(d_i|\boldsymbol{\theta}, \theta)\pi(\boldsymbol{\theta}|\varnothing)}{Z_\varnothing(d_i)},
\end{equation}
calculated for each event using some fiducial parameter prior distribution $\pi(\boldsymbol{\theta}|\varnothing)$ which does not depend on the \change{population hyperparameters}. 
Taking $n_i$ parameter posterior samples for the $i$-th event, 

\begin{align}
    \mathcal{L}_\mathrm{tot}(\vec{d}|\boldsymbol{\Lambda}) \simeq &\prod_i^N \frac{1}{P_{\mathrm{det}}(\boldsymbol{\Lambda})} \frac{Z_\varnothing(d_i)}{n_i} \sum_k^{n_i}
    \frac{\pi(\boldsymbol{\theta}^k|\boldsymbol{\Lambda})}{\pi(\boldsymbol{\theta}^k|\varnothing)},
    \label{eq:Samples}
\end{align}
where $\boldsymbol{\theta}^k$ indicates the parameters of the $k$-th sample \citep{Thrane:2018qnx,Mandel:2018mve}.

In the case where our \firstgen{} spin distribution includes the delta function at $0$, we alter this approach to account for the lack of parameter estimation samples with precisely zero component spin. 
For each event, we produce posterior samples with two fiducial priors (which are identical except for the component spins): one uniform in spin magnitude $\pi_\chi(\boldsymbol{\theta}|\varnothing)$, which enables us to sample the entire range of spins, and one where the spin is always zero $\pi_0(\boldsymbol{\theta}|\varnothing)$, which is applicable to the delta function model.

In this case, the \firstgen{} term in Eq.~\eqref{eq:Samples} becomes
\begin{align}
\label{eq:CombinedLikelihood}
    \mathcal{L}(d_i|\boldsymbol{\Lambda},\firstgen{})
    \simeq & {} \: \frac{1}{n_i}  \left[ \lambda_0 \sum_j^{n_{i,0}}
    \frac{\pi(\boldsymbol{\theta}^j|\boldsymbol{\Lambda},\firstgen)}{\pi_0(\boldsymbol{\theta}^j|\varnothing)} \right. \nonumber\\ 
    & {} \left. + (1-\lambda_0) \sum_k^{n_{i,\chi}} \frac{\pi(\boldsymbol{\theta}^j|\boldsymbol{\Lambda},\firstgen)}{\pi_\chi(\boldsymbol{\theta}^k|\varnothing)} \right]. 
\end{align}
Here, $n_{i,0}$ and $n_{i,\chi}$ are the number of samples included using the zero-spin and uniform-spin respectively, and $n_i =n_{i,0} + n_{i,\chi}$ is the total number of samples used;
the ratio of the number of zero- and uniform-spin samples is the ratio of the evidences calculated with the two priors,
\begin{equation}
    \frac{n_{i,0}}{n_{i,\chi}}=\frac{Z_0(d_i)}{Z_\chi(d_i)} = \frac{\int \mathrm{d}\boldsymbol{\theta} L(d_i|\theta)\pi_0(\boldsymbol{\theta}|\varnothing)}{\int \mathrm{d}\boldsymbol{\theta} L(d_i|\boldsymbol{\theta})\pi_\chi(\boldsymbol{\theta}|\varnothing)}.
\end{equation}
This procedure allows us to calculate the population likelihood even though the delta function and Beta distribution components of the spin model from Eq.~\eqref{eq:chi-model} have different ranges of support.

We use hierarchical Bayesian inference to construct a posterior for our \change{population hyperparameters}
\begin{equation}
    p(\boldsymbol{\Lambda}|\vec{d})=\frac{\mathcal{L}_\mathrm{hier}(\vec{d}|\boldsymbol{\Lambda})\pi(\boldsymbol{\Lambda})}{\int \mathrm{d}\boldsymbol{\Lambda} \mathcal{L}_\mathrm{hier}(\vec{d}|\boldsymbol{\Lambda})\pi(\boldsymbol{\Lambda})} ,
\end{equation}
where $\pi(\boldsymbol{\Lambda})$ is our prior for the \change{population hyperparameters}. 
With the exception of $m_\mathrm{max}$, we take this prior to be flat \citep[][]{LIGOScientific:2018jsj}.
To account for uncertainties in the location of the PISN mass gap inherent in different sets of assumptions about nuclear reaction rates, stellar rotation, accretion, and fallback \citep{Farmer:2019jed,Mapelli:2019ipt,vanSon:2020zbk}, we take a Gaussian prior on $m_\mathrm{max}$ with a mean of $50 M_{\odot}$ and standard deviation of $10 M_{\odot}$. 

We use \textsc{gwpopulation} \citep{Talbot:2019okv} and \textsc{dynesty} \citep{2020MNRAS.493.3132S} within the \textsc{Bilby} framework \citep{Ashton:2018jfp} to sample the likelihoods in Eq.~\eqref{eq:Likelihood} and Eq.~\eqref{eq:CombinedLikelihood}. 
Parameter estimation for each event is also performed using \textsc{Bilby} \citep{Romero-Shaw:2020owr}, following the settings used to produce GWTC-1 results \citep{LIGOScientific:2018mvr}. 
GW data from LIGO and Virgo are obtained from the Gravitational Wave Open Science Center \citep{Abbott:2019ebz}.

\section{Application to GWTC-1}\label{sec:GWTC-1}
\subsection{Inferred Populations}

\begin{figure}

\includegraphics[width=0.45\textwidth]{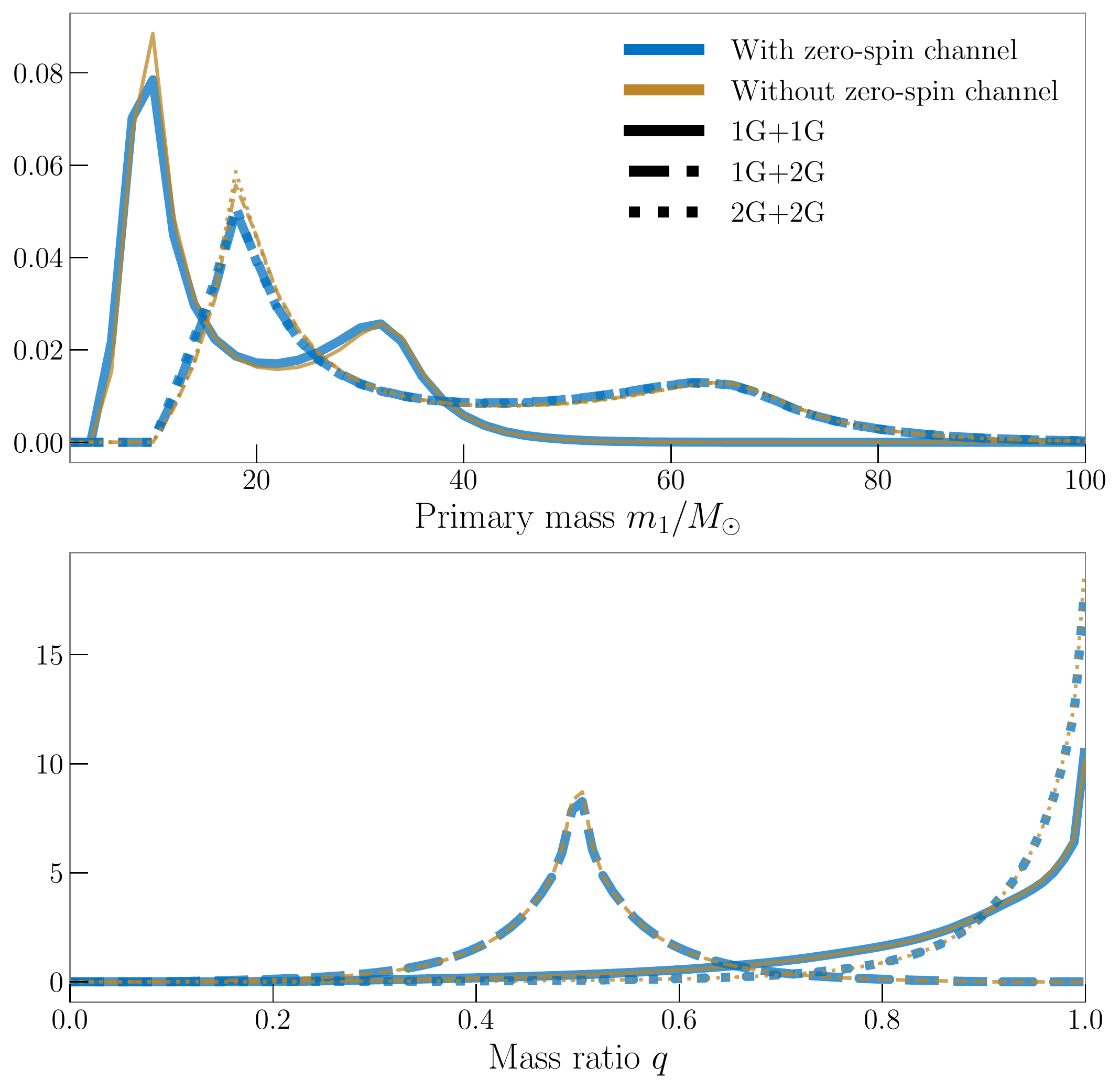}
\caption{Posterior predictive distributions for primary mass $m_1$ and mass ratio $q$. The solid, dashed, and dotted lines are the \firstgen{}, \halfgen{}, \secondgen{} distributions, respectively. The \halfgen{} and  \secondgen{} primary masses are drawn from the same distributions. In blue, we plot the distributions inferred when allowing for the zero-spin formation channel, and the distributions inferred when excluding this channel are plotted in orange. }
\label{fig:massppds}
\end{figure}

\begin{figure}

\includegraphics[width=0.45\textwidth]{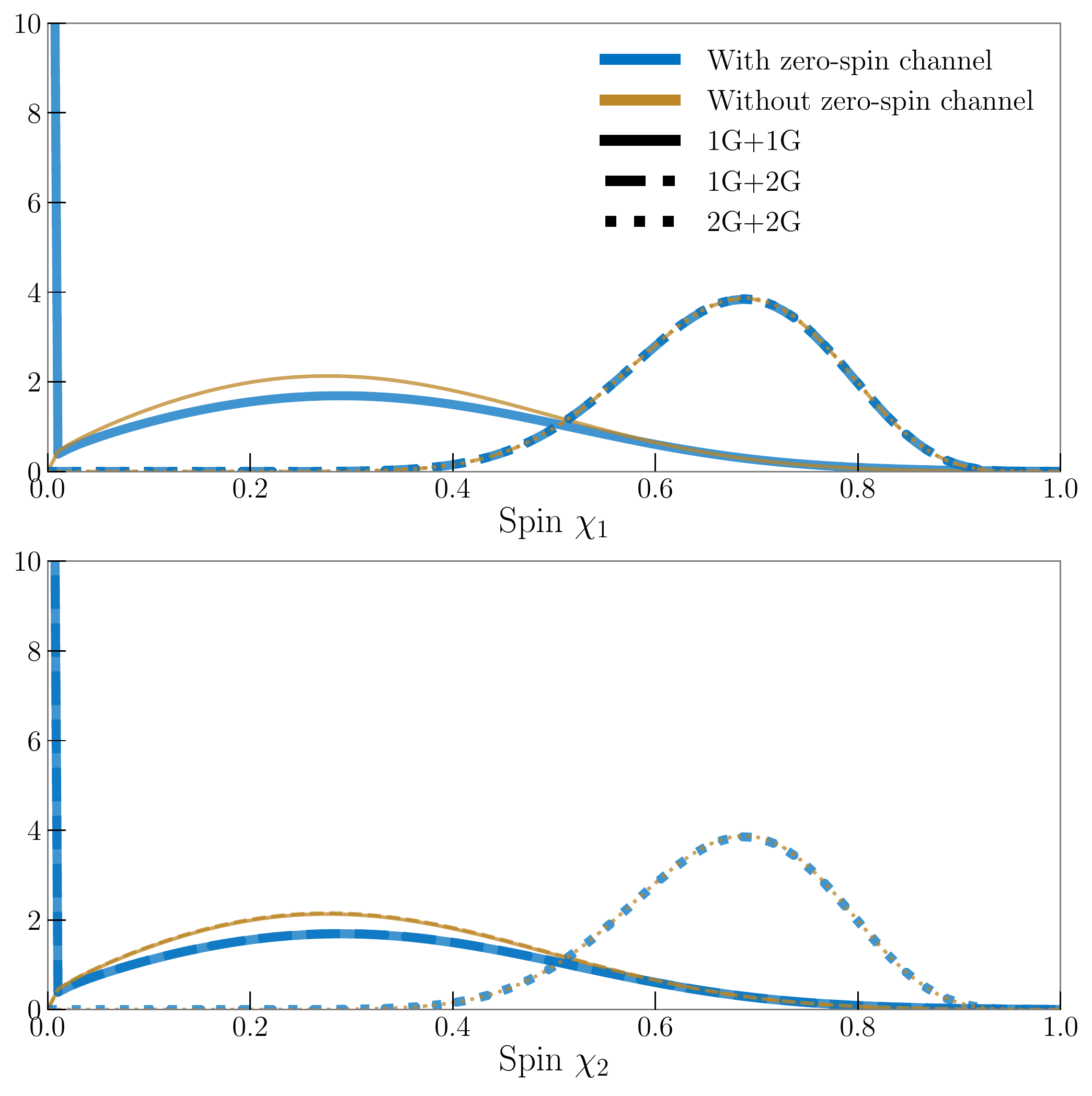}
\caption{Posterior predictive distributions for the component black hole spins. The solid, dashed, and dotted lines are the \firstgen{}, \halfgen{}, \secondgen{} distributions, respectively. The \halfgen{} and \secondgen{} primary spins are drawn from the same distributions, as are the \firstgen{} and \halfgen{} secondary spins. In blue, we plot the distributions inferred when allowing for the zero-spin formation channel, and distributions inferred when excluding this channel are plotted in orange.}
\label{fig:spinppds}

\end{figure}

\begin{figure}
\includegraphics[width=0.45\textwidth]{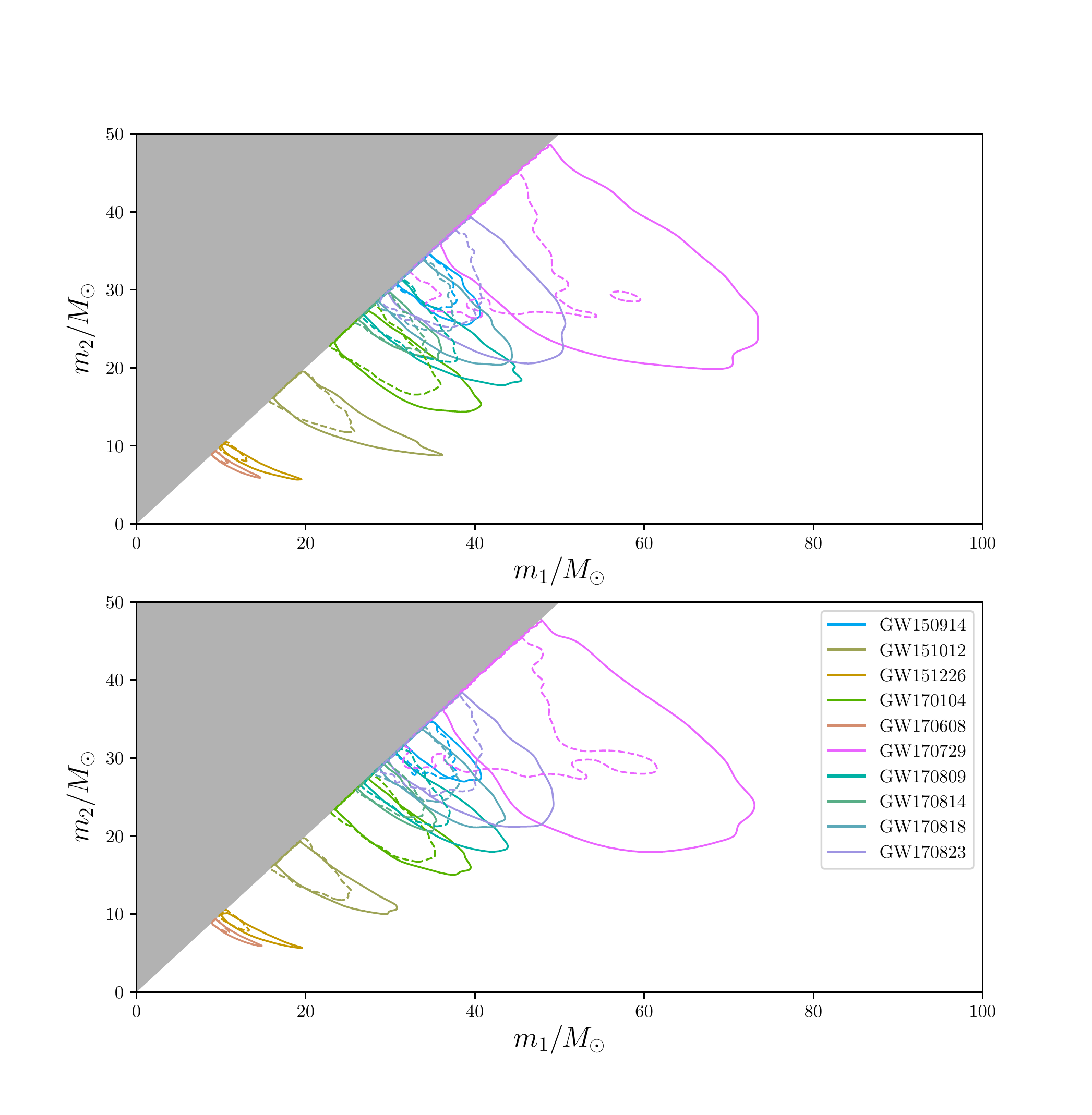}
\caption{Reweighted GWTC-1 mass posteriors using our inferred hierarchical population model as a prior. Contours indicate the $90\%$ credible areas. The original posteriors from \citet{LIGOScientific:2018mvr} are indicated with solid lines, and the reweighted posteriors are shown with dashed lines. 
\textit{Top:} results reweighted using the model inferred when excluding the zero-spin channel.
\textit{Bottom:} results reweighted using the population model inferred when allowing for the zero-spin formation channel. 
The exclusion of the zero-spin channel pushes the highest-mass events toward lower masses.
Including zero-spin allows for more retained \twoG{} black holes and hence more efficient hierarchical mergers, which, in turn, allows for larger masses.
In both cases, the region of support at high primary mass ($\sim 60 M_\odot$) in the reweighted GW170729 posterior is due to the hierarchical component of the population prior.}
\label{fig:pe}

\end{figure}

We apply the above analysis using the $10$ binary black hole observations contained in GWTC-1~\citep{LIGOScientific:2018mvr}, and infer \change{population hyperparameters} for our hierarchical model. 
\change{The inferred \change{population hyperparameters} are discussed in detail in Appendix~\ref{sec:appendix}.}
We plot the posterior predictive distributions for the \firstgen{}, \halfgen{}, and \secondgen{} populations in Fig.~\ref{fig:massppds} and Fig.~\ref{fig:spinppds}.
The \change{population hyperparameters} governing the \firstgen{} mass distribution (see Fig.~\ref{fig:massppds} in Appendix~\ref{sec:appendix}) are consistent with the results in \citet{LIGOScientific:2018jsj}. 
The Gaussian mass component corresponding to PPSN buildup is well constrained to $\mu_m \simeq \GWTCOneGaussmppLow$--$ \GWTCOneGaussmppHigh$, but we recover our prior on the location of the PISN maximum-mass cutoff $m_\mathrm{max}$. We find that $99\%$ of \firstgen{} black holes are less than $\GWTCOneGaussOneGMassULNinetyNine M_{\odot}$, in agreement with $45 M_{\odot}$ found in \citet{LIGOScientific:2018jsj}, and that $99\%$ of black holes in the combined multigeneration population are less than $\GWTCOneGaussAllMassULNinetyNine M_{\odot}$.
In Fig.~\ref{fig:spin} of Appendix~\ref{sec:appendix}, we show \change{population hyperparameters} for the \firstgen{} spin distribution. 

The fraction $\lambda_{0}$ of black holes from the zero-spin formation channel is constrained to be less than \GWTCOneGaussdeltachiULNinetyNine{} at the $99\%$ credible level, and is consistent with $\lambda_{0}=0$. 
Therefore, these GW observations suggest that at least some \firstgen{} binary black holes have spinning components, consistent with \citet{Miller:2020zox}, and not all \oneG{} black holes have extremely low ($< 0.01$) spins as would be expected if all progenitor stars had efficient angular momentum transfer \citep{Fuller:2019sxi}. 
We find that $90\%$ of \firstgen{} primary black holes have a spin magnitude less than $\GWTCOneGaussOneGSpinULNinety$.

In the bottom panel of Figure~\ref{fig:pe}, we reweight the GWTC-1 mass posteriors to apply our inferred hierarchical population model as a prior: the primary effect acts to constrain the mass ratio compared to the fiducial prior used in the initial parameter inference. 
Upon reweighting, the $90\%$ credible interval on the primary black hole mass for GW170729 becomes $35$--$55$ $M_{\odot}$, compared to $40$--$66$ $M_{\odot}$ with the default prior \citep{LIGOScientific:2018mvr}. 

\subsection{Relative merger rates}

\begin{figure}
\includegraphics[width=0.45\textwidth]{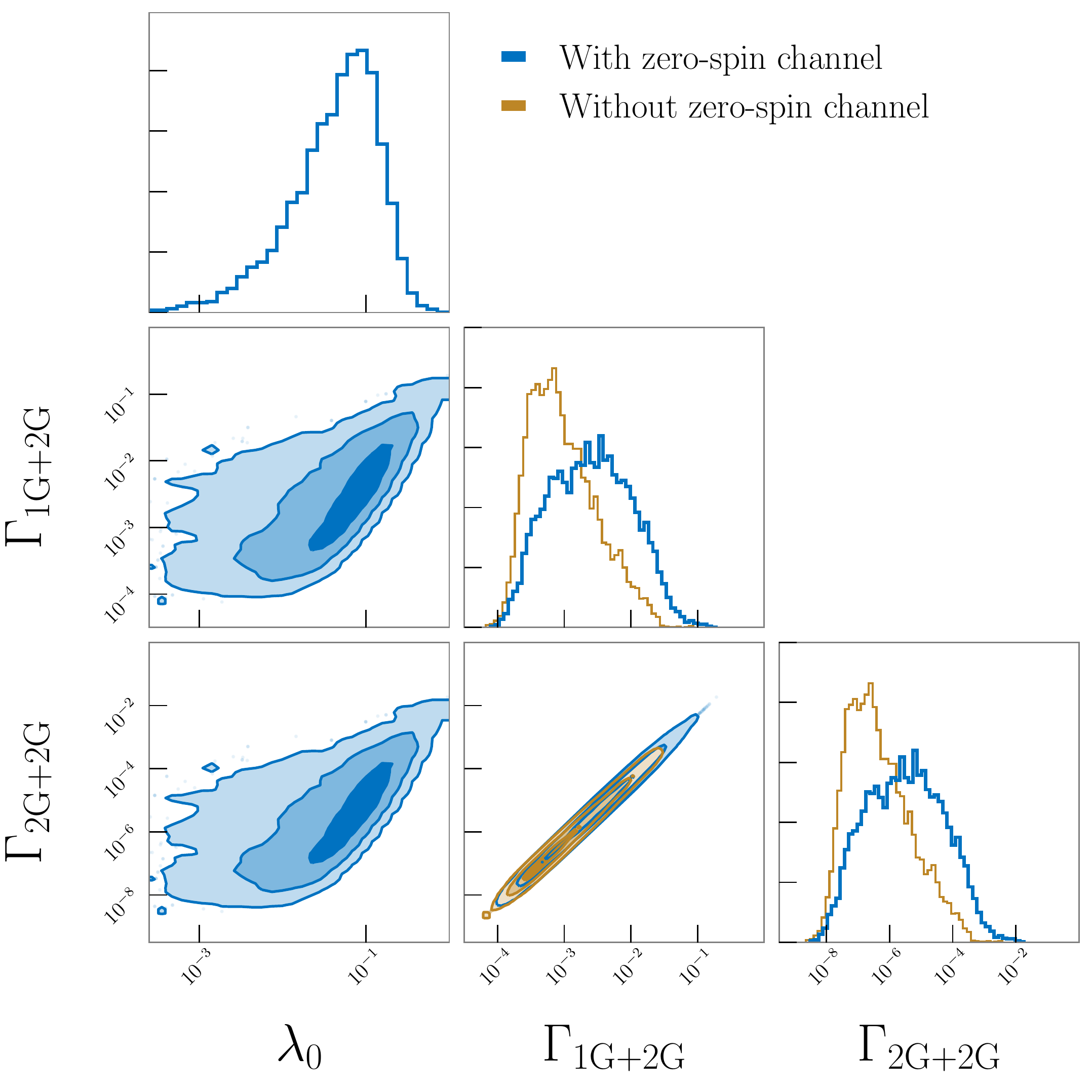}
\caption{Posteriors of the inferred branching ratios, which are the relative \halfgen{} vs. \firstgen{} and \secondgen{} vs. \firstgen{} merger rates, and the fraction of \firstgen{} binary black holes with zero-spin $\lambda_0$. In blue we plot the results when we allow for the zero-spin formation channel and in orange we plot the results when excluding the zero-spin formation channel (fixing $\lambda_0 = 0$). }
\label{fig:branching}
\end{figure}

As shown in Fig.~\ref{fig:branching}, we find that the relative rates $\Gamma_\halfgen$ and $\Gamma_\secondgen$ are strongly correlated with the fraction $\lambda_{0}$ of \oneG{} black holes that form in the zero-spin channel. 
These branching ratios are set by the fraction of \firstgen{} merger products that are retained in a typical cluster. 
Since merging binaries with non-spinning components experience lower recoil  velocities than those with non-negligible spin, the inclusion of the zero-spin formation channel drastically affects the retention fraction \change{in the globular cluster potential}, and consequently the branching ratios. 

We find the median relative rates $\Gamma_\halfgen$ and $\Gamma_\secondgen$ to be $\GWTCOneGaussBranchingHalfMed$ and $\GWTCOneGaussBranchingTwoMed$, respectively, with $99\%$ upper limits of $\GWTCOneGaussBranchingHalfULNinetyNine$ and $\GWTCOneGaussBranchingTwoULNinetyNine$. 
Adopting a fiducial binary black hole merger rate of $\sim 50~\mathrm{Gpc^{-3}\,yr^{-1}}$ \citep{LIGOScientific:2018jsj} as a \firstgen{} merger rate (we do not explicitly infer the rate as part of our model) these $99\%$ upper limits would imply merger rates of $\lesssim 2.5~\mathrm{Gpc^{-3}\,yr^{-1}}$ for $\Gamma_\halfgen$ and $\lesssim 0.06~\mathrm{Gpc^{-3}\,yr^{-1}}$ for $\Gamma_\secondgen$.
Rerunning our analysis without the zero-spin subcomponent, the median branching ratios $\Gamma_\halfgen{}$ and $\Gamma_\secondgen{}$ become $\GWTCOneNoZeroSpinGaussBranchingHalfMed$ and $\GWTCOneNoZeroSpinGaussBranchingTwoMed$, respectively, with $99\%$ upper limits of $\GWTCOneNoZeroSpinGaussBranchingHalfULNinetyNine$ and $\GWTCOneNoZeroSpinGaussBranchingTwoULNinetyNine$. 

As the rates are much lower, we are less likely to observe hierarchical mergers than when there are black holes with effectively zero spin. 
The sensitivity of the merger rates to spin could potentially enable us to place tight constraints on the spins of \oneG{} black holes---which are difficult to measure directly from GW observations \citep{Poisson:1995ef,Purrer:2015nkh,Vitale:2014mka,LIGOScientific:2018mvr}---through the constraints on the hierarchical merger rate. 

The lower branching ratios inferred when excluding the zero-spin formation channel affect the shape of the overall multigenerational population, with little support for primary masses in the PPSN mass gap. 
In the top panel of Fig.~\ref{fig:pe}, we plot the reweighted component mass posterior samples for the $10$ events in GWTC-1, with the population model excluding the zero-spin component as a prior. 
The reduced hierarchical merger rates lead to smaller support for masses above the upper mass cutoff, and the $90\%$ interval on the primary black hole mass for GW170729 tightens to $34$--$53 M_{\odot}$. 
Without the zero-spin population subcomponent, the $90\%$ upper limit on \firstgen{} primary black hole spin magnitude becomes $\GWTCOneNoZeroSpinGaussOneGSpinULNinety$. 

The inferred branching ratios are consistent with Monte Carlo modeling of binary black hole populations in globular clusters; in the most extreme case where all black holes are assumed to be born with zero spin, such modeling predicts $\approx 13\%$ ($1\%$) of merging binary black holes are \halfgen{} (\secondgen{}) systems \citep{Rodriguez:2019huv}. 
As the natal spins of black holes increase, the retention fractions and relative rates precipitously drop as the recoil kicks become stronger. 
\citet{Rodriguez:2019huv} find that if black hole natal spins are assumed to be $\chi = 0.5$, that the number of black holes with a \twoG{} component drops to $\lesssim 1\%$ of the total population. 


\subsection{Odds ratios for the hierarchical merger scenario}

\begin{figure}
\includegraphics[width=0.45\textwidth]{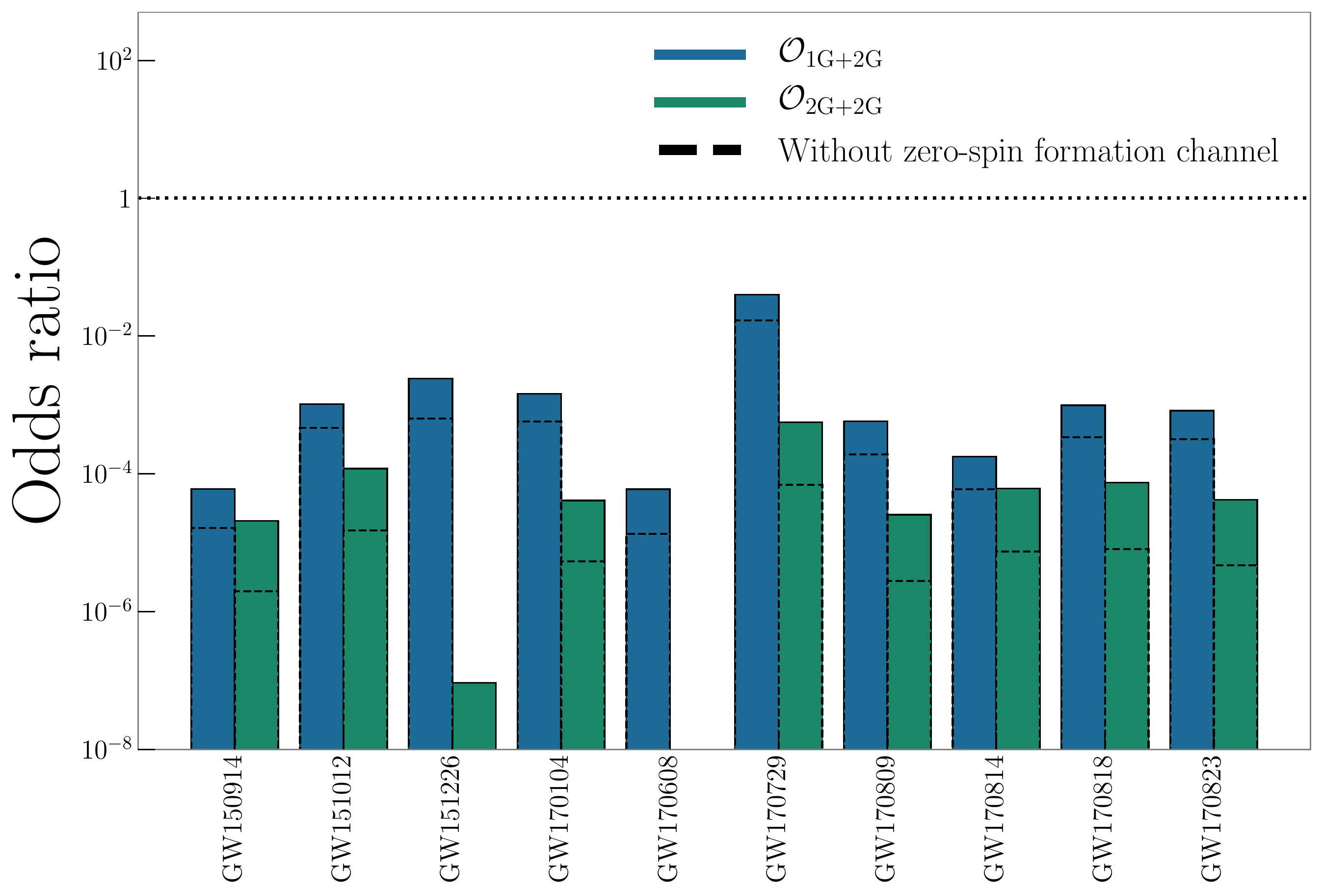}
\caption{Hierarchical/\firstgen{} odds ratios for each of the GWTC-1 events. The odds for \halfgen{} origin are plotted in blue, while the odds for \secondgen{} origin are in green. The dashed lines indicate the odds when we use the model inferred when excluding the zero-spin channel. The dotted line indicates even odds.}
\label{fig:odds}
\end{figure}

With our multigenerational model, we also can calculate the hierarchical/\firstgen{} odds ratio $\mathcal{O}$ for each event. 
If the parameter distributions of each generational subpopulation were known, the odds ratio that the $i$-th observation came from a \halfgen{} system versus a \firstgen{} system would be
\begin{align}
    \mathcal{O}^i_\halfgen{} \equiv & {}\: \frac{P(\halfgen|d_i)}{P(\firstgen|d_i)} \nonumber\\
    = & {}\: \frac{Z(d_i|\halfgen)}{Z(d_i|\firstgen)}\frac{P(\halfgen{})}{P(\firstgen{})}, \label{eq:exactOdds}
\end{align}
where the first term in Eq.~\eqref{eq:exactOdds} is the ratio of evidences for the observation given the \halfgen{} and \firstgen{} subpopulations \citep[a Bayes factor;][]{Kimball:2019mfs}, and the second term is the prior odds (relative rates) of mergers of the two generations. 
However, as we do not know the exact form of the underlying population, our uncertainty in the \change{population hyperparameters} affects both the relative rates and the ratio of evidences.
To take this into account, we marginalize over the \change{population hyperparameters}, weighting by our posterior probability distribution $p(\boldsymbol{\Lambda}|\vec{d})$, yielding
\begin{align}
    \mathcal{O}^i_\halfgen{} = \frac{\int  \mathrm{d} \boldsymbol{\Lambda} Z(d_i|{\boldsymbol{\Lambda}}, \halfgen) \zeta_\halfgen{}(\boldsymbol{\Lambda}) p(\boldsymbol{\Lambda}|\vec{d})}{\int \mathrm{d} \boldsymbol{\Lambda} Z(d_i|{\boldsymbol{\Lambda}}, \firstgen) \zeta_\firstgen{}(\boldsymbol{\Lambda}) p(\boldsymbol{\Lambda}|\vec{d})}.
    \label{eq:odds}
\end{align}
Here, the evidence for the \halfgen{} population is
\begin{equation}
    Z(d_i|{\boldsymbol{\Lambda}}, \halfgen) = \int \mathrm{d}\boldsymbol{\theta}  L(d_i|\boldsymbol{\theta})\pi(\boldsymbol{\theta}|{\boldsymbol{\Lambda}}, \halfgen),
\end{equation}
while the \firstgen{} evidence $Z(d_i|{\boldsymbol{\Lambda}}, \firstgen)$ is defined similarly, and $\zeta_\halfgen{}$ and $\zeta_\firstgen{}$ are the hierarchical merger fractions.
The odds ratio for a \secondgen{} system versus a \firstgen{} system $\mathcal{O}^i_\secondgen{}$ can be calculated by swapping \halfgen{} to \secondgen{} in Eq.~\eqref{eq:odds}.

We calculate these odds ratios for all $10$ events in GWTC-1 \citep{LIGOScientific:2018mvr}, and plot the results in Fig.~\ref{fig:odds}. 
For GW170729---the event with the most massive primary black hole---we favor a \firstgen{} over a \halfgen{} origin with $25$:$1$ odds when including the zero-spin formation channel.
The probability that GW170729 is of hierarchical origin (either \halfgen{} or \secondgen{}) is $4\%$. 
GW151226, which has the most confidently measured non-zero spin \citep{Abbott:2016nmj,LIGOScientific:2018mvr,Miller:2020zox}---we find that $\log_{10} (Z_\chi/Z_0) = 6.5$---has the second highest probability ($0.2\%$) for a hierarchical origin.
Across all $10$ systems in GWTC-1, we find the probability that at least one binary black hole system is of hierarchical origin is $5\%$. 

As the inferred branching ratios are much smaller when excluding the zero-spin formation channel, the odds ratios for hierarchical origin \change{in our globular cluster model}, shown in dashed lines in Fig.~\ref{fig:odds}, are reduced by $\sim3$--$5$. 
If we exclude the zero-spin channel, we find that GW170729 most likely has a \firstgen{} origin by a factor of $60$:$1$, and that the probability of at least one event being of hierarchical origin is only $2\%$.

The branching ratios are also dependent on the escape velocity of the dynamical environment. 
If we increase our cluster mass to $10^8$ $M_{\odot}$, typical of a nuclear star cluster, the branching ratios---and hence the odds ratios in favor of a hierarchical origin---increase by $\sim 1$--$3$ orders of magnitude. 
Since our transfer functions for \halfgen{} and \secondgen{} populations are tuned to globular cluster simulations, a robust analysis of an nuclear star cluster hierarchical merger scenario would require more detailed study.

To check how our prior on $m_\mathrm{max}$ affects our results we rerun the analysis with a uniform prior between $20 M_\odot$ and $200 M_{\odot}$. 
While we infer a peak in the posterior on $m_\mathrm{max}$ near $40 M_{\odot}$, we find support all the way out to $200 M_{\odot}$, well above any of the GWTC-1 black hole masses, indicating that we are insensitive to the existence of the mass gap (discussed further in Appendix~\ref{sec:appendix}). 
The odds ratios in favor of the GWTC-1 events being hierarchical mergers remains largely the same, with a small increase in favor of hierarchical mergers as the prior for $m_\mathrm{max}$ extends down to $20 M_\odot$.
With this prior, the GW170729 \halfgen{} odds ratio is $0.041$. 
Allowing $m_\mathrm{max}$ to extend to larger values makes it easier to incorporate high-mass systems into the \firstgen{} population. 
Cutting on the maximum of the $m_\mathrm{max}$ prior from $200 M_{\odot}$ to $40 M_{\odot}$ increases the GW170729 \halfgen{} odds ratio to $0.046$. 
Overall, our conclusions are not significantly affected by the prior assumptions on $m_\mathrm{max}$ as none of the systems lack posterior support for having masses below the PISN mass gap.

\section{Conclusions}\label{sec:conclusions}

GW observations have demonstrated that binary black holes merge to form more massive black holes \citep{Abbott:2016blz}. 
If these merger products form a new binary, they may again become a GW source. 
The complete catalog of GW sources may therefore contain a mixture of \oneG{} black holes formed from stellar collapse, and \twoG{} black holes formed in mergers. 
In using the population of GW sources to infer the formation mechanisms for black holes, e.g., if their progenitors are subject to PPSN, it is necessary to account for the potential presence of \twoG{} black holes to prevent our conclusions being biased.
However, it is difficult to concretely distinguish between \oneG{} and \twoG{} black holes, as the populations overlap in properties. 
We perform an analysis that self-consistently infers both the fraction of binaries containing \twoG{} black holes, and the fundamental properties of the population of \firstgen{} binaries.

Our analysis uses phenomenological models to describe the binary black hole population. 
The models are calibrated to reproduce the features seen in simulations of globular clusters \citep{Rodriguez:2019huv}. 
The fraction of \twoG{} black holes that are retained in a cluster following a merger depends sensitively upon the spins of \oneG{} black holes, as larger spins results in larger GW recoil kicks. 
Simulations of massive stars with efficient angular momentum transfer predict that black holes would form with spins $\lesssim 0.01$ \citep{Fuller:2019sxi}. 
Therefore, our population model also includes the possibility of a fraction of \oneG{} black holes that have effectively zero spin.
Our analysis demonstrates that this is a potentially key ingredient in the search for hierarchical mergers.

We apply our approach to the $10$ binary black holes found by LIGO and Virgo in their first two observing runs \citep{LIGOScientific:2018mvr}. 
We find that: 
\begin{enumerate}
    \item The \firstgen{} population is fit by a steep power law with exponent $\alpha > \GWTCOneGaussalphaLow$ plus a Gaussian component with mean $\mu_m = \GWTCOneGaussmppMed^{+\GWTCOneGaussmppPlus}_{-\GWTCOneGaussmppMinus} M_\odot$. 
    We find an upper cutoff to the power law of $m_\mathrm{max} = \GWTCOneGaussmmaxMed^{+\GWTCOneGaussmmaxPlus}_{-\GWTCOneGaussmmaxMinus} M_\odot$, but this is dominated by our choice of prior. 
    Across the multigenerational population, we find that $99\%$ of black holes in binaries have masses $m_1 \lesssim \GWTCOneGaussAllMassULNinetyNine M_\odot$.
    Overall, the \firstgen{} population is consistent with the mass distributions inferred in \citet{LIGOScientific:2018jsj}.
    \item The fraction of \firstgen{} binaries with zero spin is $\lambda_0 < \GWTCOneGaussdeltachiULNinetyNine$ with $99\%$ probability, and $90\%$ of \firstgen{} primary black holes have spins less than $\GWTCOneGaussOneGSpinULNinety$. Excluding the zero-spin formation channel, $90\%$ of \firstgen{} primary black holes have spins less than $\GWTCOneNoZeroSpinGaussOneGSpinULNinety$
    \item The median merger rates of \halfgen{} and \secondgen{} binaries relative to \firstgen{} binaries are inferred to be $\GWTCOneGaussBranchingHalfMed$ and $\GWTCOneGaussBranchingTwoMed$, respectively, with $99\%$ upper limits of $\GWTCOneGaussBranchingHalfULNinetyNine$ and $\GWTCOneGaussBranchingTwoULNinetyNine$.
    The relative rates are tightly correlated with the fraction of \oneG{} black holes with zero spin. 
    Excluding the zero-spin subcomponent of our spin distribution, the relative rates drop to $\GWTCOneNoZeroSpinGaussBranchingHalfMed$ and $\GWTCOneNoZeroSpinGaussBranchingTwoMed$ respectively, with $99\%$ upper limits of $\GWTCOneNoZeroSpinGaussBranchingHalfULNinetyNine$ and $\GWTCOneNoZeroSpinGaussBranchingTwoULNinetyNine$.
    Since the relative rates and spins are tightly linked, a measurement of one would pin down the other.
    \item The $10$ binary black holes from GWTC-1 are all consistent with being \firstgen. 
    Given the rarity of \halfgen{} and \secondgen{} mergers, this is not surprising. GW170729's source, which is the most massive of the observed systems, is still found to most likely have a first-generation origin. This result is not especially sensitive to the allowed range for $m_\mathrm{max}$, as the masses for GW170729 are consistent with being below the PISN gap.
\end{enumerate}
We cannot make a definite conclusion about the presence of hierarchical mergers amongst this catalog of $10$ events. 

The analysis is currently limited to considering binary black holes formed in globular clusters. 
In reality, we expect that binary black holes form in other environments as well. 
Black holes in the field are unlikely to undergo a hierarchical merger. 
On the other hand, those formed in a nuclear star cluster are much more likely to be retained and available to form hierarchical mergers due to their higher escape velocities \citep{Antonini:2016gqe,Antonini:2018auk,Yang:2019cbr}. 

Including alternative channels is necessary for definitively identifying hierarchical mergers, as this and other evolutionary channels, such as stellar collisions in young stellar clusters \citep{DiCarlo:2019fcq}, growth in active galactic nucleus disks \citep{McKernan:2012rf}, or consecutive mergers in quadruple systems \citep{Fragione:2020aki}, can grow black holes to masses above the PISN cutoff. 
The rate at which these mass-gap black holes form merging binaries is highly uncertain.
If these black holes merge, they would be (incorrectly) classified as hierarchical mergers within our globular cluster picture.

Our method can be extended to include additional subpopulations. 
This would require defining new models, for example, including an aligned-spin distribution, as detailed in Eq.~\eqref{eq:iso-spin}, to model binaries formed via isolated evolution \citep{Kalogera:1999tq,Rodriguez:2016vmx}. 
Including more subpopulations adds parameters to the likelihood, Eq.~\eqref{eq:overall-single-sub}. 
With only the $10$ binaries, a relatively simple model is prudent \citep{LIGOScientific:2018jsj}.
However, this will change as the catalog grows with further observing runs \citep{Vitale:2015tea,Stevenson:2017dlk,Zevin:2017evb,Talbot:2017yur}.

The third observing run of LIGO and Virgo began in April 2019 and was suspended in March 2020. 
The first binary black hole detection of the third observing run has recently been announced: GW190412 \citep{LIGOScientific:2020stg}, a system with an unequal component masses; in Appendix~\ref{sec:GW190412}, we examine how adding this new event to the GWTC-1 updates our results, again finding that all binaries are consistent with being \firstgen{}.\footnote{\change{The recently announced GW190814 has an uncertain nature and could be a binary black hole or a neutron star--black hole binary \citep{Abbott:2020khf}, and we do not consider it in our analysis.}} 
Full results from the third observing run are still to be announced. 
The fourth observing run, which will extend the global GW detector network to include KAGRA \citep{Akutsu:2018axf}, is scheduled to start in mid 2021 \citep{Aasi:2013wya}. 
As we gather more observing time, and improve the sensitivity of the detector network, we expect the number of observations and the rate of discoveries to increase. 
With larger catalogs of events it will be possible to make more precise measurements of the population, and we will be able to determine whether hierarchical mergers play a significant role in the GW catalog.
Furthermore, improvements in the detectors' low-frequency sensitivity will improve their ability to detect higher-mass binaries \citep{Abbott:2017iws}. 
The next generation of ground-based detectors offers the opportunity to perform the same measurements across cosmic time \citep{Kalogera:2019sui}. 
With the precise population measurements coming from larger catalogs we can infer the details of the physical processes that shape black hole formation; however, for these conclusions to be accurate, it is necessary to account for the population being a mix of both \oneG{} black holes and the products of mergers.

\acknowledgements{}
The authors thank Kyle Kremer, Carl Rodriguez, Tom Dent, Mario Spera, and Zoheyr Doctor for their expert advice in constructing this study. 
The authors are grateful to Riccardo Buscicchio and Ethan Payne for their careful comments on the analysis; Reed Essick for help on calculating detector sensitivities, and Scotty Coughlin for high-performance computing support. 
This research has made use of data obtained from the Gravitational Wave Open Science Center (\href{https://www.gw-openscience.org}{www.gw-openscience.org}), a service of LIGO Laboratory, the LIGO Scientific Collaboration and the Virgo Collaboration. LIGO is funded by the US National Science Foundation (NSF). Virgo is funded by the French Centre National de Recherche Scientifique (CNRS), the Italian Istituto Nazionale della Fisica Nucleare (INFN) and the Dutch Nikhef, with contributions by Polish and Hungarian institutes. 
This work is supported by the NSF Grant PHY-1607709 and through the Australian Research Council (ARC) Centre of Excellence CE170100004. 
C.K. is supported supported by the National Science Foundation under grant DGE-1450006.
C.P.L.B. is supported by the CIERA Board of Visitors Research Professorship. 
E.T. is supported through ARC Future Fellowship FT150100281 and CE170100004.
This research was supported in part through the computational resources from the Grail computing cluster at Northwestern University---funded through NSF PHY-1726951---and staff contributions provided for the Quest high performance computing facility at Northwestern University, which is jointly supported by the Office of the Provost, the Office for Research, and Northwestern University Information Technology. 
The authors are grateful for computational resources provided by the LIGO Laboratory and supported by NSF Grants PHY-0757058 and PHY-0823459.
This document has been assigned LIGO document number \href{https://dcc.ligo.org/LIGO-P2000131/public}{LIGO-P2000131}.

\appendix

\section{GWTC-1 hyperparameter distributions}
\label{sec:appendix}

\begin{figure*}
\includegraphics[width=\textwidth]{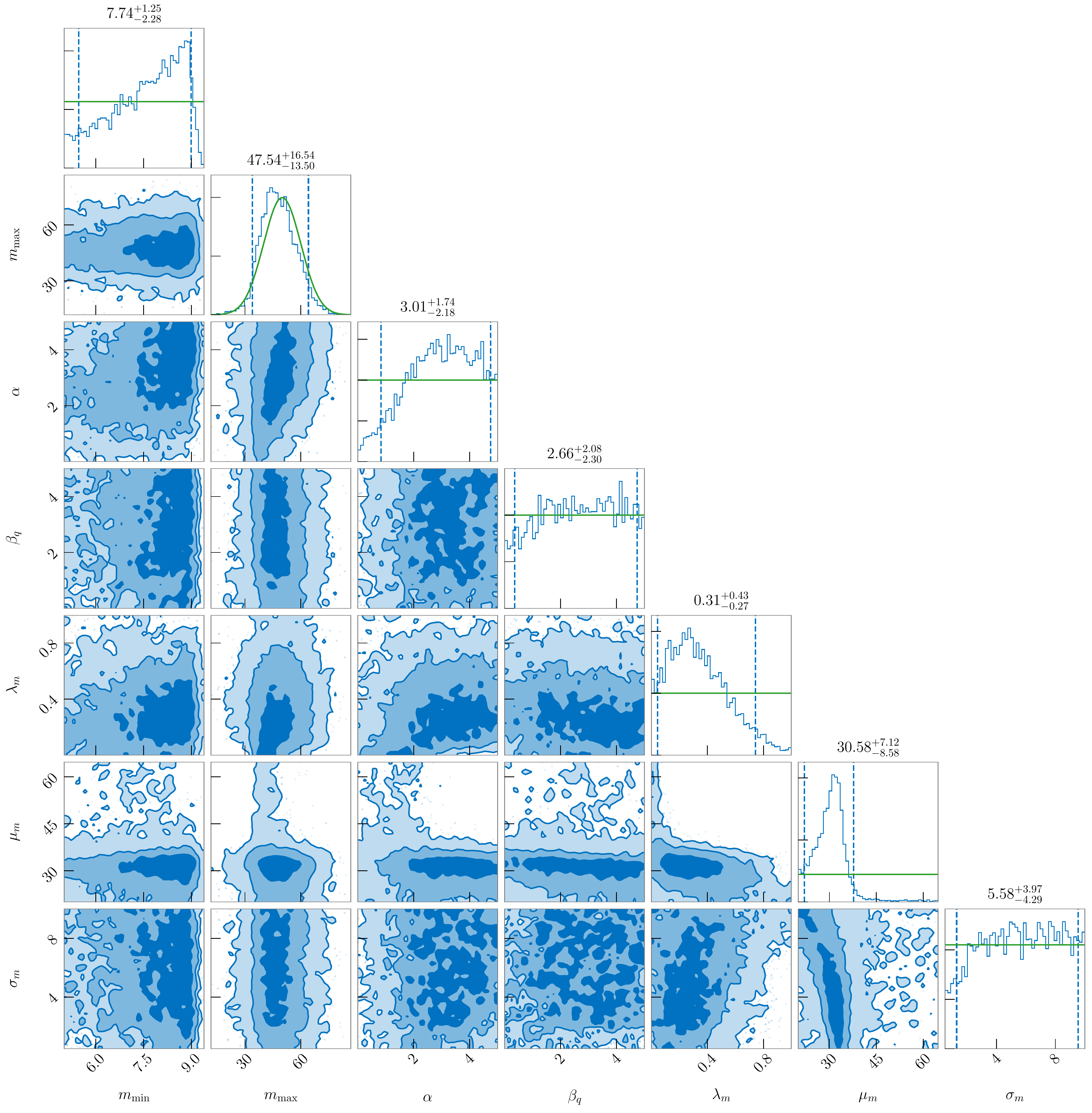}
\caption{Posterior distributions of the \change{population hyperparameters} governing the mass and mass ratio distributions, when we allow a fraction of \oneG{} black holes to form in the zero-spin channel. The dashed lines give the $90\%$ credible intervals, and the green lines indicate the priors.}
\label{fig:mass}
\end{figure*}

\begin{figure*}
\includegraphics[width=\textwidth]{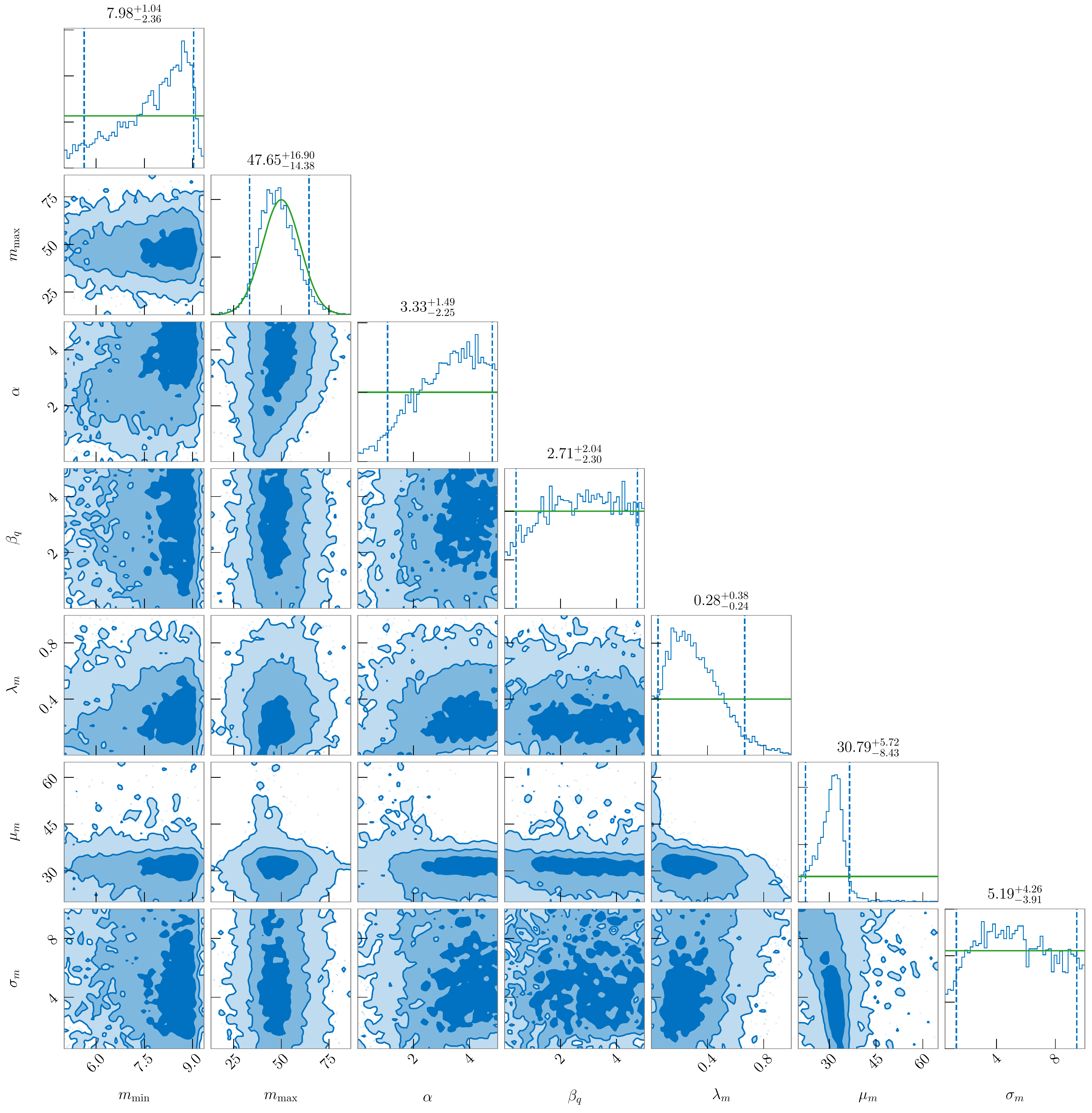}
\caption{Inferred posterior distributions of the \change{population hyperparameters} governing the mass and mass ratio distributions when excluding the zero-spin formation channel. The dashed lines give the $90\%$ credible intervals intervals, and the green lines indicate the priors.}
\label{fig:massnozero}
\end{figure*}

\begin{figure}
\includegraphics[width=0.9\textwidth]{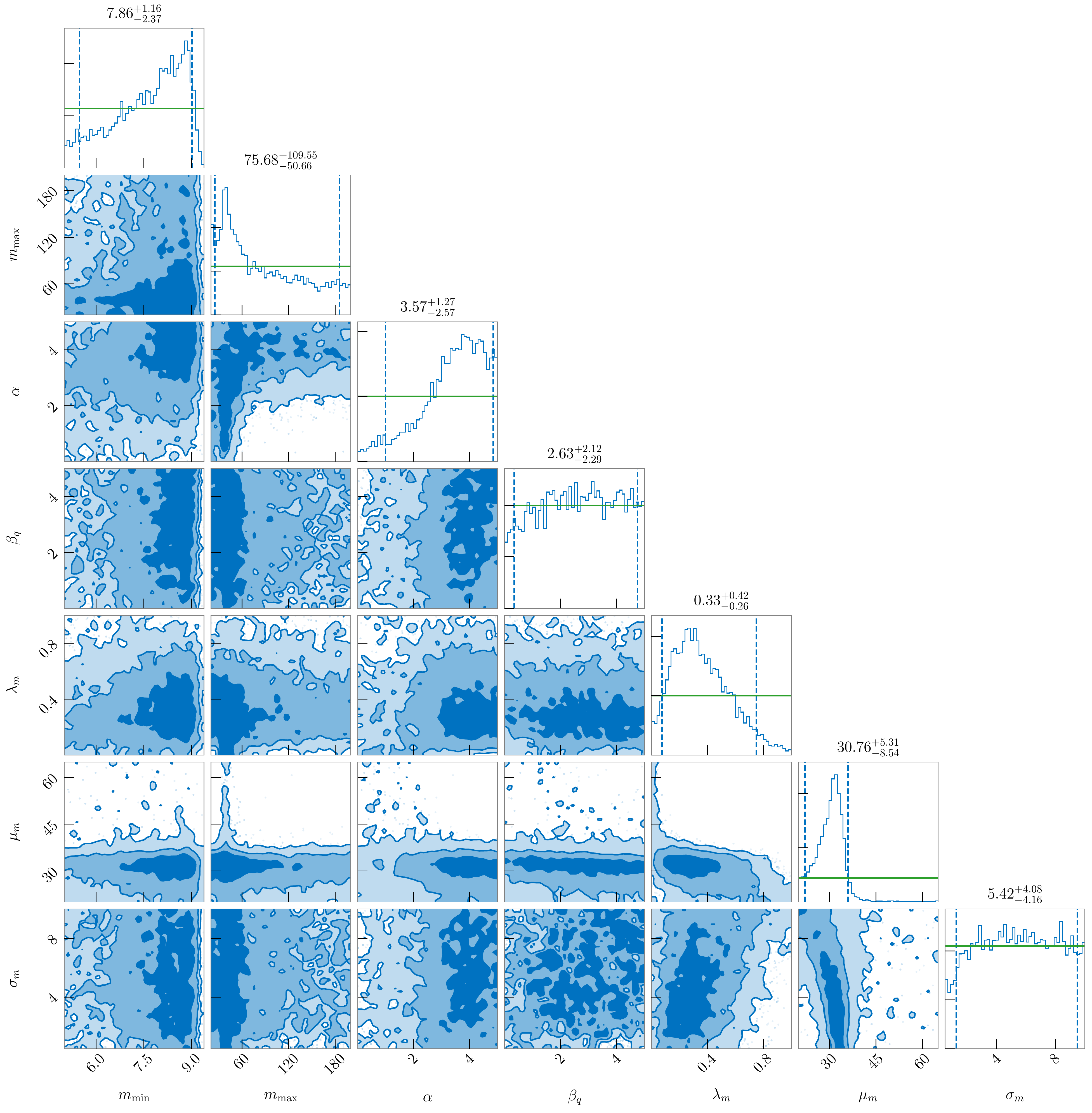}
\caption{Posterior distributions of the \change{population hyperparameters} governing the mass distributions when we assume a flat prior on $m_\mathrm{max}$ The dashed lines give the $90\%$ credible intervals intervals, and the green lines indicate the priors.}
\label{fig:massflat}
\end{figure}

Here we present the full sets of inferred population parameter $\boldsymbol{\Lambda}$ posteriors for our population models.
In Fig.~\ref{fig:mass}, we plot the parameters determining the mass distributions, as defined in Eq.~\eqref{eq:m1-model} and Eq.~\eqref{eq:q-model}, for our default model.
In Fig.~\ref{fig:massnozero} we plot the equivalent mass \change{population hyperparameters} for the model excluding the zero-spin subcomponent, and in Fig.~\ref{fig:massflat} we plot the mass \change{population hyperparameters} when we switch to using a uniform prior for $m_\mathrm{max}$. 
The results are largely consistent between model choices.

When using the astrophysically motivated prior for $m_\mathrm{max}$, the posterior closely follows the prior. 
The posterior on $m_\mathrm{max}$ is more restricted at smaller values of the power-law index $\alpha$: when the mass distribution is flatter we are more sensitive to the upper cutoff than when the distribution sharply decreases with mass and we can increase the upper cutoff with little consequence \citep{Fishbach:2017zga}. 
When switching to the uniform prior on $m_\mathrm{max}$ we see the same qualitative behavior with varying $\alpha$. 
For steep power laws ($\alpha \gtrsim 2$), we are effectively insensitive to the existence of an upper cutoff, but for flatter power laws ($\alpha \lesssim 1$), the dearth of higher-mass black holes means that there is little posterior support for $m_\mathrm{max} \gtrsim 45 M_\odot$. 

The power-law index $\alpha$ has more support for higher ($\alpha \gtrsim 2$) values. 
Our posterior on $\alpha$ is truncated by our choice of prior.
\citet{LIGOScientific:2018jsj} found that the posterior on $\alpha$ becomes uninformative at large values ($\alpha \gtrsim 4$), with all values matching equally well.

The Gaussian component of the mass spectrum has a mean well constrained between $\mu_m \simeq \GWTCOneGaussmppLow$--$\GWTCOneGaussmppHigh M_\odot$. 
The exception to this is when $\lambda_m \sim 0$, as then the Gaussian component is negligible and so can be positioned anywhere. 
There is a correlation between the width of the Gaussian component $\sigma_m$ and the mean \citep{Talbot:2018cva}, with smaller $\mu_m$ permissible when $\sigma_m$ is larger, as this enables the upper edge of the Gaussian to stay in place. 
The value of $\sigma_m$ is not well constrained by the current set of observations.

The posteriors for the minimum mass $m_\mathrm{min}$ are largely unconstrained. 
As the GW detectors are less sensitive to low-mass systems, it is more difficult to place constraints on this end of the distribution \citep{Fishbach:2017zga,Talbot:2018cva,LIGOScientific:2018jsj}. 
The lower limit of the $m_\mathrm{min}$ distribution is set by our prior, and the upper limit is set by the least massive black hole amongst our observations. 

The mass ratio is degenerate with the spin \citep{Poisson:1995ef,Baird:2012cu,Farr:2015lna}. 
Fixing spins to be zero breaks the mass--spin degeneracy results in a more equal mass ratio and a larger $m_2$ for a system of a given chirp mass. 
However, the inclusion of the zero-spin subcomponent makes little difference for our inferred mass ratio distribution, with the posterior for the power-law index $\beta_q$ being largely determined by our assumed prior.

\begin{figure}
\includegraphics[width=0.45\textwidth]{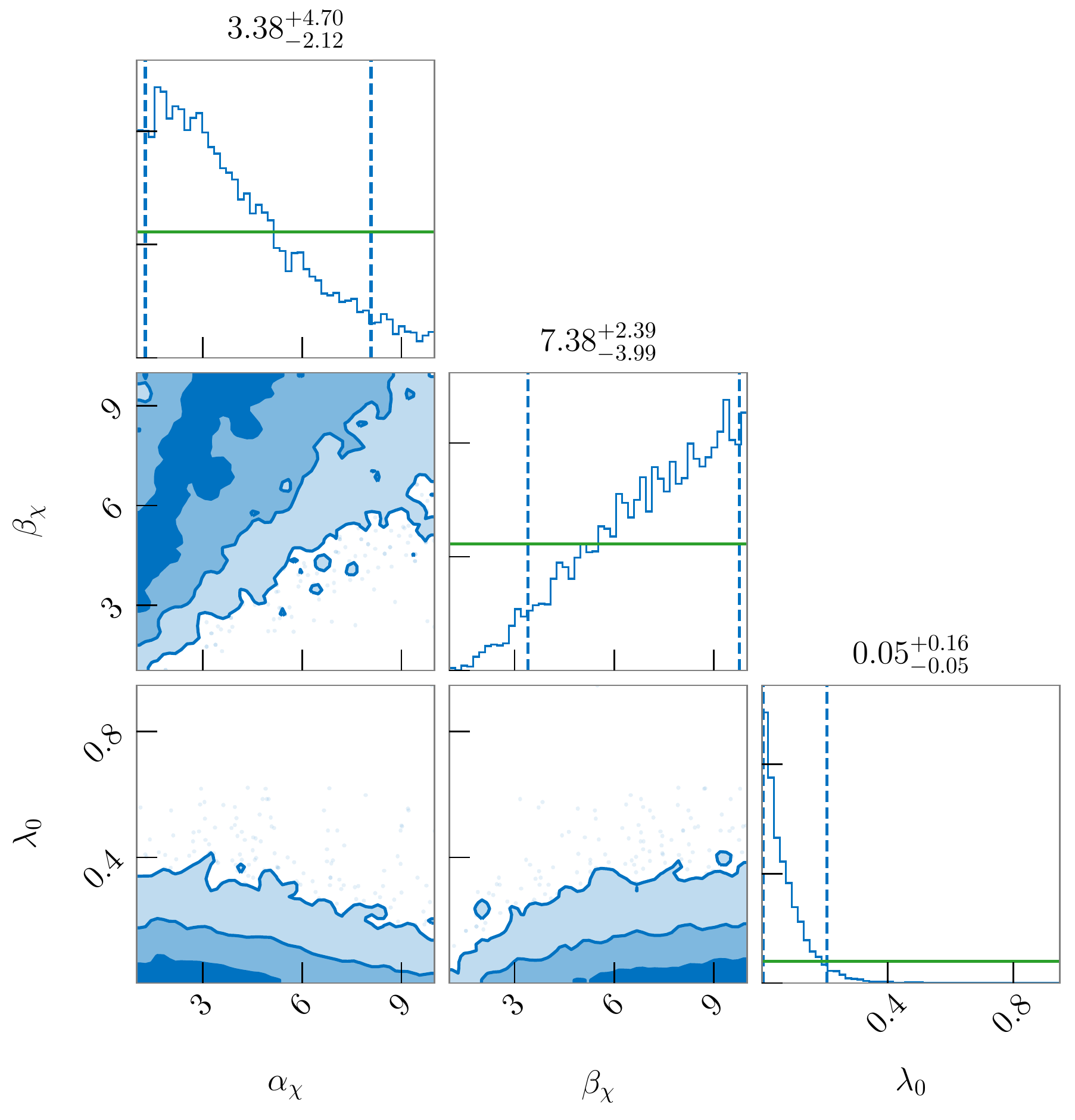}
\caption{Posterior distributions of the \change{population hyperparameters} governing the spin distributions, when we allow a fraction of \oneG{} black holes to form in the zero-spin channel. The dashed lines give the $90\%$ credible intervals intervals, and the green lines indicate the priors.}
\label{fig:spin}
\end{figure}

\begin{figure}
\includegraphics[width=0.45\textwidth]{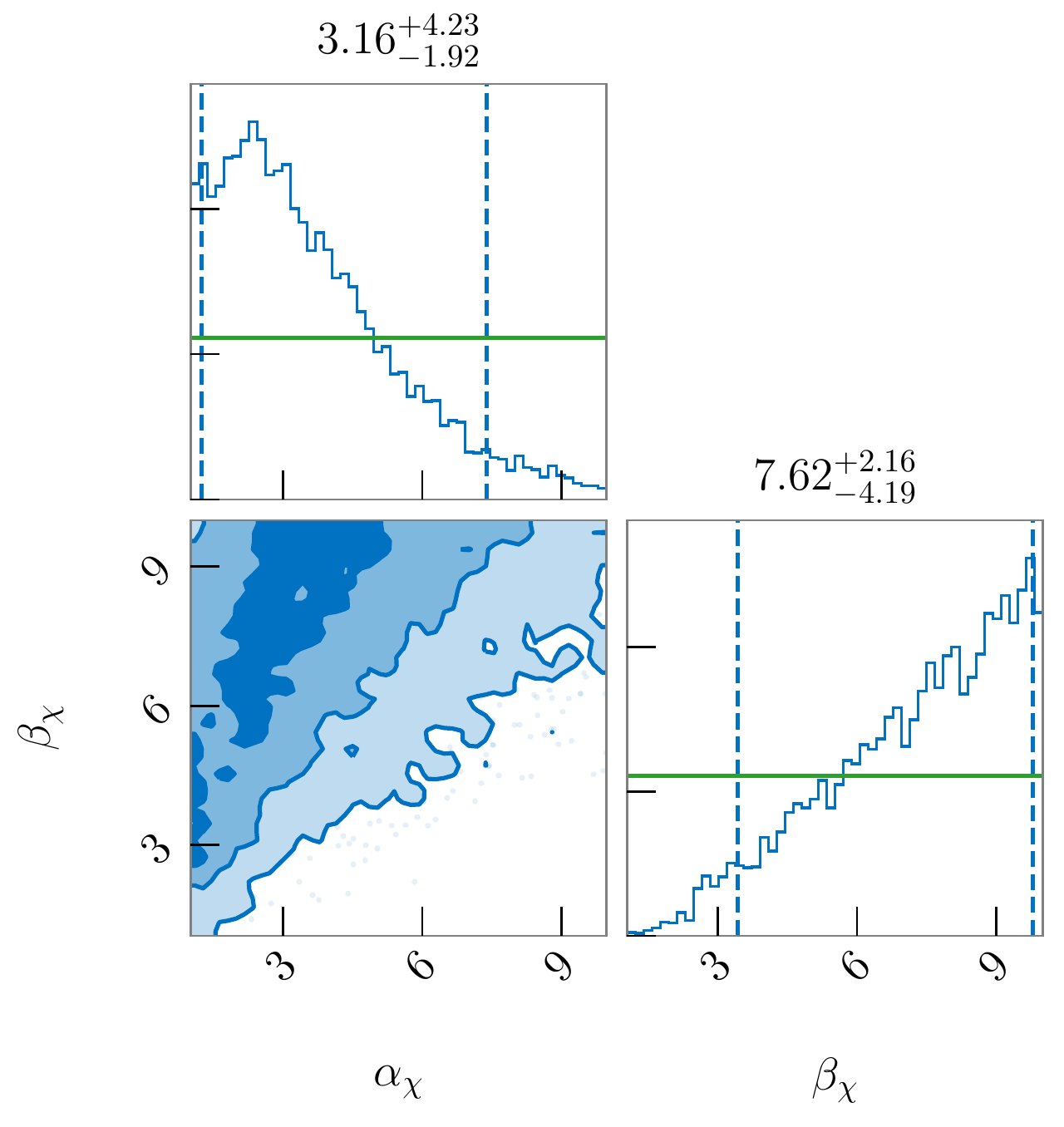}
\caption{Posterior distributions of the \change{population hyperparameters} governing the spin distributions when excluding the zero-spin formation channel. The dashed lines give the $90\%$ credible intervals intervals, and the green lines indicate the priors.}
\label{fig:spinnozero}
\end{figure}

\begin{figure}
\includegraphics[width=0.45\textwidth]{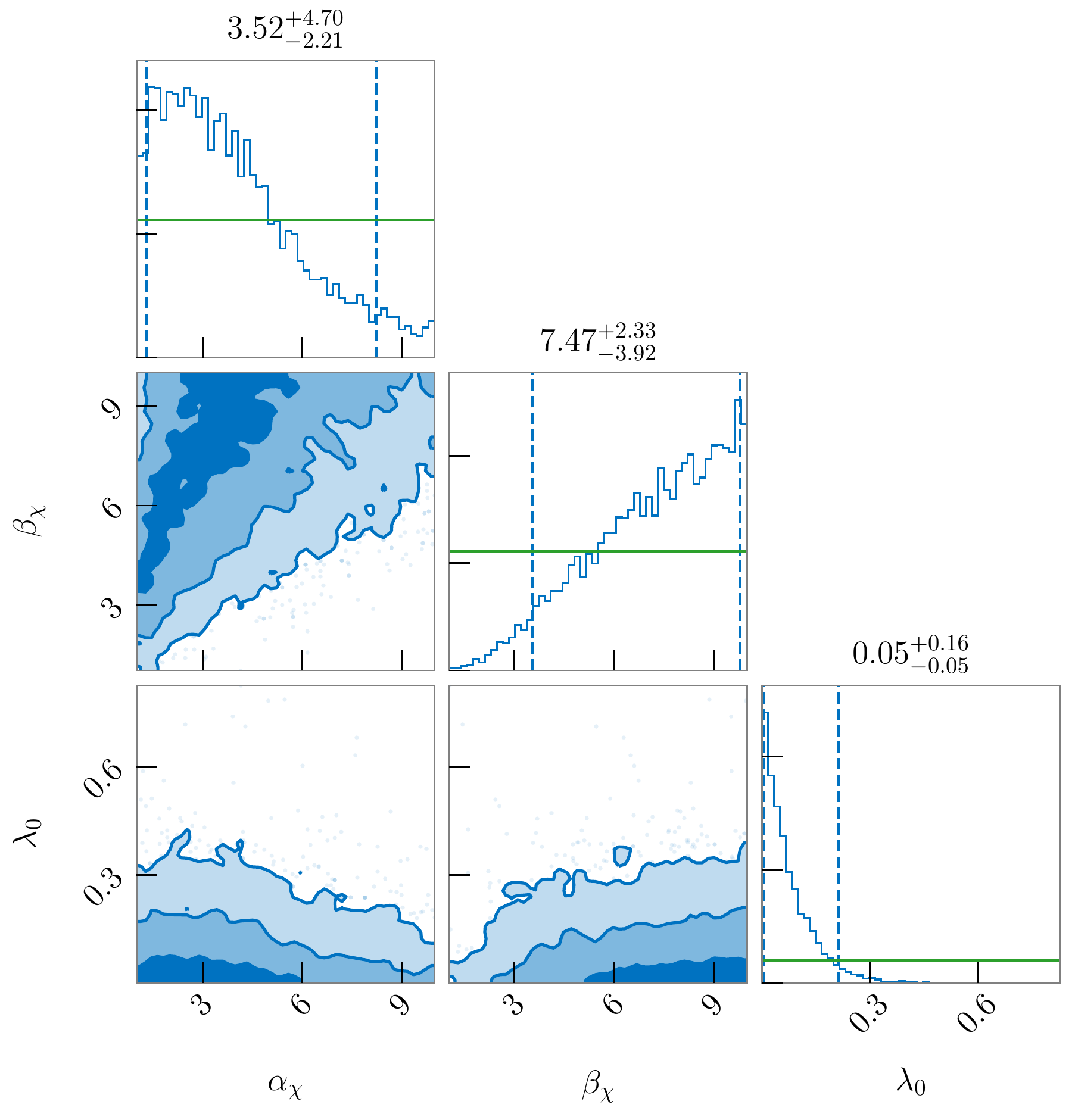}
\caption{Posterior distributions of the \change{population hyperparameters} governing the spin distributions when we assume a flat prior on $m_\mathrm{max}$ The dashed lines give the $90\%$ credible intervals intervals, and the green lines indicate the priors.}
\label{fig:spinflat}
\end{figure}

In Fig.~\ref{fig:spin}, we plot the parameters determining the spin distributions, as defined in Eq.~\eqref{eq:chi-model}, for our default model.
In Fig.~\ref{fig:spinnozero} we plot the equivalent spin \change{population hyperparameters} for the model with $\lambda_0 = 0$, and in Fig.~\ref{fig:spinflat} we plot the spin \change{population hyperparameters} when using a uniform prior for $m_\mathrm{max}$. 
The $m_\mathrm{max}$ prior makes negligible difference to the spin distributions. 
There is no simple correlation between the fraction of \firstgen{} binaries with zero spin $\lambda_0$ and the other \change{population hyperparameters}. 
The $\lambda_0$ distribution is peaked at $0$ and shows that many \firstgen{} binaries are not well described by both black holes having near-zero spins. 
In all cases we favor models with $\alpha_\chi < \beta_\chi$, which corresponds to distributions which decrease with increasing spin magnitude \citep{Farr:2017uvj,LIGOScientific:2018jsj}.


\section{Including GW190412}
\label{sec:GW190412}

GW190412 is the first announced binary black hole detection of the third observing run of LIGO and Virgo \citep{LIGOScientific:2020stg}. 
It is exceptional on account of its mass ratio, which is inferred as $q = 0.28^{+0.13}_{-0.06}$ assuming the fiducial parameter estimation prior. 
The large difference in component masses would not be surprising for a hierarchical merger \citep{Rodriguez:2020viw,Gerosa:2020bjb}, so here we investigate how our results change including GW190412. 
For this, we use parameter-estimation results for GW190412 from \citet{Zevin:2020gxf}.
Since GW190412 has been especially selected for publication, we cannot assume that using GWTC-1 plus GW190412 is a fair representation of the binary black hole population, and so results using these $11$ systems should be considered as preliminary, pending the completion of the catalog from the third observing run.

\begin{figure}
\includegraphics[width=0.9\textwidth]{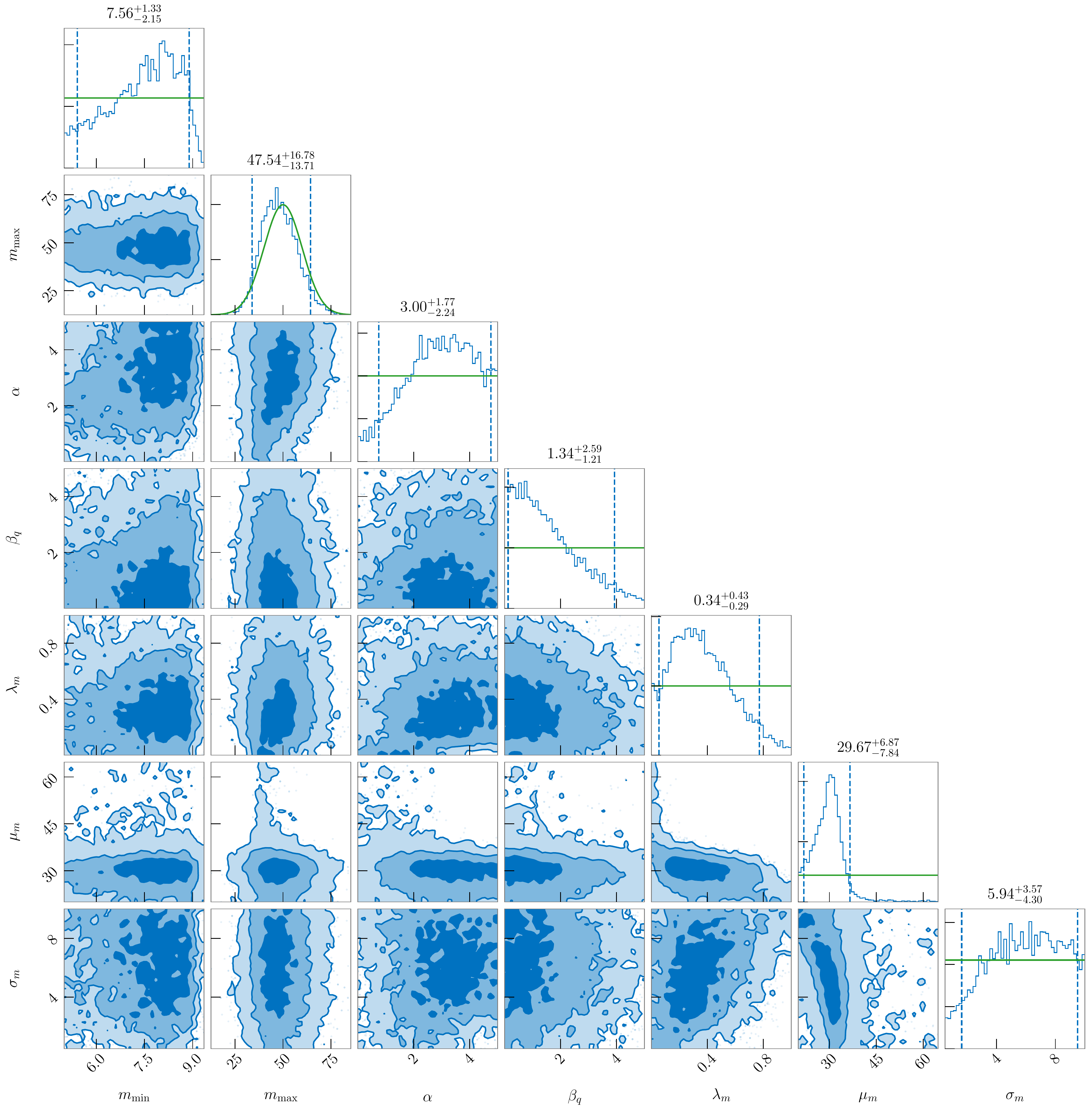}
\caption{Posterior distributions of the \change{population hyperparameters} governing the mass and mass ratio distributions when we include GW190412, inferred with our model that allows a fraction of \oneG{} black holes to form in the zero-spin channel. The dashed lines give the $90\%$ credible intervals, and the green lines indicate the priors.}
\label{fig:GW190412Mass}
\end{figure}

\begin{figure}
\includegraphics[width=0.45\textwidth]{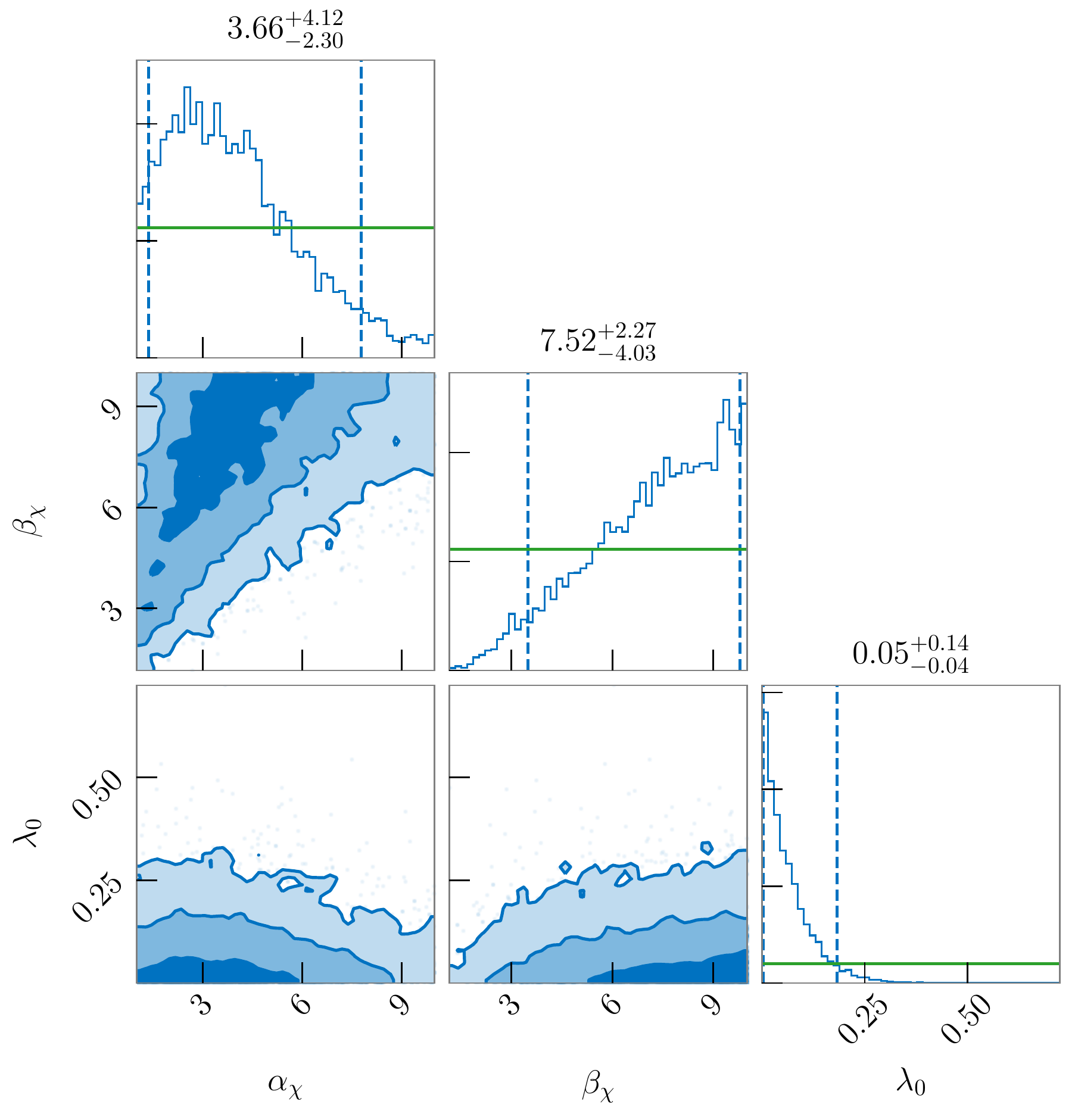}
\caption{Posterior distributions of the \change{population hyperparameters} governing the spin distributions when we include GW190412, and allow a fraction of \oneG{} black holes to form in the zero-spin channel. The dashed lines give the $90\%$ credible intervals intervals, and the green lines indicate the priors.}
\label{fig:GW190412Spin}
\end{figure}

In Fig.~\ref{fig:GW190412Mass}, we plot the \change{population hyperparameters} for the mass and mass ratio distributions. 
As in \citet{LIGOScientific:2020stg}, we find that including GW190412 leads to tighter constraints on the mass ratio distribution. 
This single additional event acts as a lever arm, constraining $\beta_q$ to smaller values, flattening our inferred mass ratio distribution. 
The inferred spin parameters, shown in Fig.~\ref{fig:GW190412Spin}, are unaffected. 

\begin{figure}
\includegraphics[width=0.45\textwidth]{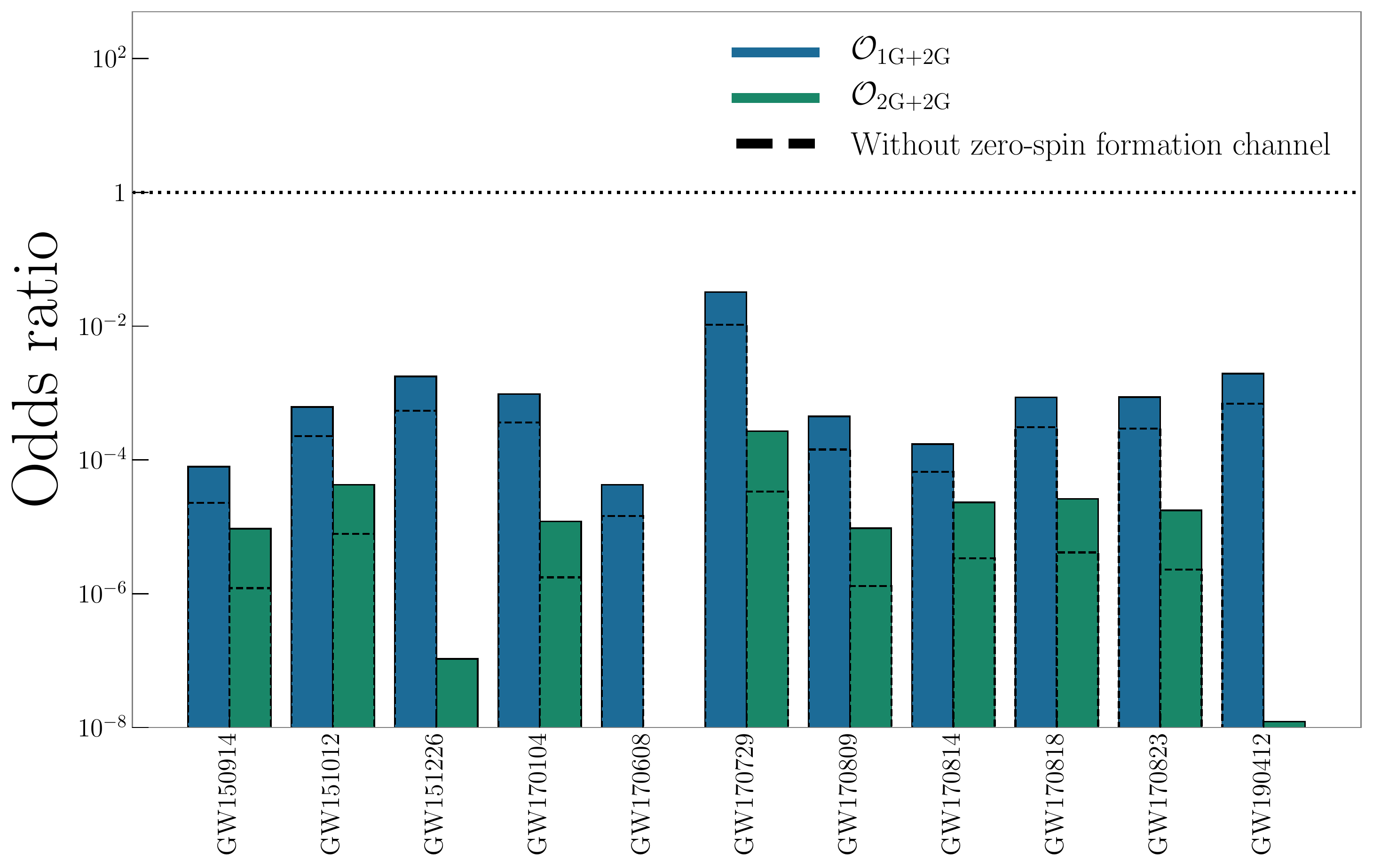}
\caption{Hierarchical/\firstgen{} odds ratios for each of the GWTC-1 events as well as GW190412. The odds for a \halfgen{} origin are plotted in blue, while the odds for a \secondgen{} origin are in green. The dashed lines indicate the odds when we use the model inferred when excluding the zero-spin channel. The dotted line indicates even odds.}
\label{fig:GW190412Odds}
\end{figure}

In Fig.~\ref{fig:GW190412Odds}, we show the odds ratios for our events having a hierarchical versus \firstgen{} origin. The extreme mass ratio of GW190412 is well explained by the \halfgen{} population, but its primary component's spin is below the \twoG{} black hole spin distribution. 
Overall, we find that GW190412 most likely has a \firstgen{} origin, at odds of $\sim 500$:$1$. 
Including GW190412 also reduces the odds of GW170729 having a \halfgen{} origin by $\sim 20\%$, since our \firstgen{} mass ratio distribution flattens and has increased support at more unequal mass ratios.  
When we increase our cluster mass to $10^8 M_{\odot}$, chosen to be typical of a nuclear star cluster, we still find that GW190412 most likely has a \firstgen{} origin, but at lower odds of $\sim 6$:$1$.


\bibliography{nextgen.bib}

\end{document}